\documentclass{article}

\usepackage{PRIMEarxiv}

\usepackage[utf8]{inputenc} 
\usepackage[T1]{fontenc}    
\usepackage{hyperref}       
\usepackage{url}            
\usepackage{booktabs}       
\usepackage{amsfonts}       
\usepackage{nicefrac}       
\usepackage{microtype}      
\usepackage{lipsum}
\usepackage{fancyhdr}       
\usepackage{graphicx}       
\graphicspath{{media/}}     

\usepackage{multirow}
\usepackage{threeparttable}
\usepackage{makecell}
\usepackage{colortbl}
\usepackage{caption}
\usepackage{subcaption}
\usepackage{xspace}
\usepackage{xcolor}
\usepackage{enumitem}
\usepackage{wrapfig}
\usepackage{booktabs}
\usepackage{tabularx}
\usepackage{tcolorbox}

\usepackage{amssymb}
\usepackage{hyperref}
\usepackage[linesnumbered,ruled,vlined]{algorithm2e}
\usepackage{amsmath}
\usepackage{wrapfig}
\usepackage[ruled,vlined]{algorithm2e}
\usepackage{url}
\definecolor{customred}{HTML}{FF5314}
\definecolor{customgreen}{HTML}{7CBB00}

\title{Can LLMs Deobfuscate Binary Code? A Systematic Analysis of Large Language Models into Pseudocode Deobfuscation}

\author{
Li Hu$^{1}$ \quad
Xiuwei Shang$^{1,2}$\thanks{Corresponding author.} \quad
Jieke Shi$^{2}$ \quad
Shaoyin Cheng$^{1}$\footnotemark[1] \quad 
Junqi Zhang$^{1}$ \quad
Gangyang Li$^{1}$ \quad \\
\textbf{Zhou Yang$^{3}$} \quad
\textbf{Weiming Zhang$^{1}$} \quad
\textbf{David Lo$^{2}$} \\
\\
$^{1}$University of Science and Technology of China, Hefei, China \\
$^{2}$Singapore Management University, Singapore \\
$^{3}$University of Alberta, Edmonton, Canada \\
\\
\texttt{pdxbshx@mail.ustc.edu.cn, shangxw@mail.ustc.edu.cn, xwshang@smu.edu.sg} \\
\texttt{jiekeshi@smu.edu.sg, sycheng@ustc.edu.cn, jqzh@ustc.edu.cn, ligangyang@mail.ustc.edu.cn,} \\
\texttt{zy25@ualberta.ca, zhangwm@ustc.edu.cn, davidlo@smu.edu.sg} \\
}

\begin{document}
\maketitle

\newcommand{\sysname}{{\sc BinDeObfBench}\xspace}

\begin{abstract}

Deobfuscating binary code remains a fundamental challenge in reverse engineering, as obfuscation is widely used to hinder analysis and conceal program logic. Although large language models (LLMs) have shown promise in recovering semantics from obfuscated binaries, a systematic evaluation of their effectiveness is still lacking. In this work, we present \sysname, the first comprehensive benchmark for assessing LLM-based binary deobfuscation across diverse transformations spanning pre-compilation, compile-time, and post-compilation stages. Our evaluation shows that deobfuscation performance depends more on reasoning capability and domain expertise than on model scale, and that task-specific supervised fine-tuning consistently outperforms broad domain pre-training. Reasoning models can maintain robustness under severe obfuscation, generalize across different instruction set architectures (ISAs) and optimization levels. In-context learning benefits standard models but yields limited gains for reasoning models. Overall, our study highlights the importance of task-specific fine-tuning and reasoning-driven strategies, and positions \sysname as a basis for future work in binary deobfuscation.

\end{abstract}

\keywords{Reverse Engineering, Binary Code Deobfuscation, Benchmarking, Large Language Models}

\section{Introduction}

Binary code obfuscation is a widely used technique that applies semantics-preserving transformations to increase structural complexity, thereby deliberately impeding the analysis of the resulting executable binaries~\cite{heffner2004obfuscation} and obscuring sensitive implementation details to thwart reverse engineering and protect intellectual property~\cite{schrittwieser2016protecting}. However, obfuscation also be exploited by attackers to evade detection by security systems~\cite{li2022chosen}, concealing the unauthorized use of code~\cite{ko2017coat} and facilitating the development of malware~\cite{malware}. Therefore, binary code deobfuscation is essential for both security analysis and software maintenance, enabling analysts to understand program logic and detect hidden threats when source code is unavailable~\cite{jiang2024binaryai}.

\begin{figure}[t]
  \centering
  \includegraphics[width=0.92\linewidth]{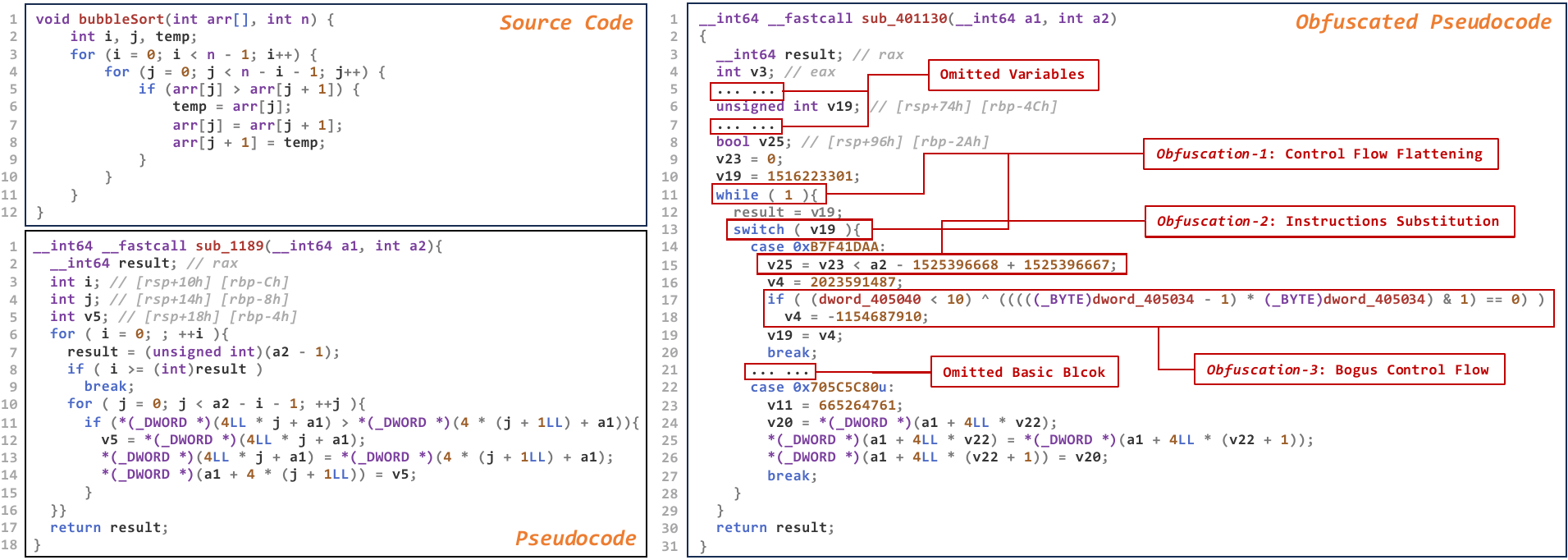}
  \caption{Example of a bubble sort function in three representations: (top-left) source code, (bottom-left) pseudocode decompiled by IDA Pro, (right) obfuscated pseudocode with combined obfuscation transformations.}
  \vspace{-3.3ex}
  \label{fig:example}
\end{figure}

Despite its importance, binary code deobfuscation remains challenging in practice. In real-world scenarios, the production of obfuscated binaries, whether for protection or by adversaries, often involves combining multiple obfuscation techniques with additional custom and undocumented transformations. \autoref{fig:example} shows an example in which a compact nested-loop implementation of bubble sort is transformed into a flattened switch-based control flow with injected bogus branches. These combined transformations substantially increase program complexity and hinder semantic understanding and recovery, which poses challenges to existing deobfuscation approaches such as \cite{lee2024poster, lee2023simplifying, yadegari2015generic, kochberger2021sok}. Most of these approaches rely on manually crafted rules and heuristic-based techniques, and are often tailored to specific obfuscation patterns, resulting in limited generalizability and poor adaptability to diverse obfuscation schemes \cite{mariano2024control,liu2021mba}.

Recently, Large Language Models (LLMs) have demonstrated strong capabilities across software engineering tasks, including code generation~\cite{fakhoury2024llm}, automatic repair \cite{xia2023automated}, and code completion~\cite{zhang2024llm}. Their performance in security-critical domains, such as binary analysis, has also been impressive, with studies~\cite{hu2024degpt, shang2024far, shang2025binmetric, tan2024llm4decompile} showing that LLMs generalize well to tasks including function naming~\cite{jiang2025beyond}, summary generation~\cite{zhu2025misum, jin2023binary}, and type inference~\cite{wang2025typeforge}. Even without explicit structural information, their semantic modeling and code understanding abilities allow LLMs to uncover latent patterns in binary code. Despite these strengths, there is still a lack of systematic evaluation of the performance of LLMs in binary code deobfuscation, which is a substantially more challenging task that requires recovering complete program logic from binaries.

In this paper, we propose \sysname, the first benchmark designed to evaluate the deobfuscation capabilities of LLMs on binary code. It comprises 2,108,736 uniquely obfuscated programs with ground truth, generated from source code through six common obfuscation transformations applied at three stages: pre-compilation, compile-time, and post-compilation. The benchmark captures the diversity of real-world software through variations in combinations of obfuscation transformations, ISAs, and optimization levels. We also design four evaluation metrics to comprehensively assess the effectiveness of deobfuscation methods in terms of lexical consistency, semantic preservation, code simplicity, and code readability. On \sysname, we evaluate nine representative models spanning three categories: (1) general-purpose or code-specific LLMs including Qwen2.5-Coder~\cite{hui2024qwen2}, Qwen3~\cite{qwen3technicalreport}, CodeLlama \cite{codellama}, Llama-3.1~\cite{llama3.1}, DeepSeek-V3~\cite{liu2024deepseek}, GPT-4o~\cite{gpt-4o}; (2) reasoning models, namely OpenAI-o1~\cite{opeai-o1} and DeepSeek-R1~\cite{guo2025deepseek}; and (3) one domain-specfic expert model tailored for binary analysis tasks, Recopilt~\cite{chen2025recopilot}. We further include an existing LLM-based approach, ChatDEOB~\cite{choi2024chatdeob}, and two traditional deobfuscation tools, D810~\cite{D810} and GooMBA~\cite{GooMBA}, for comparative evaluation. We additionally examine the impact of various factors (e.g., in-context learning examples) on LLM deobfuscation performance, and evaluate the applicability of LLM-based deobfuscators in realistic settings using maliciously obfuscated binaries.

Our evaluation identifies five key empirical findings. First, results challenge the traditional scaling assumption in binary analysis, demonstrating that reasoning capability and domain expertise outweigh raw model scale, with task-specific fine-tuning proving more effective than broad domain pre-training. Second, as obfuscation intensity increases, reasoning capability proves more critical than domain expertise, namely, surface-level knowledge alone is insufficient for complex transformations, whereas reasoning-oriented models remain more robust under severe conditions (e.g., at Level-6 obfuscation, DeepSeek-R1 preserves 62.89\% semantic fidelity, compared to 58.30\% for ChatDEOB and 54.62\% for ReCopilot). Third, ISAs and optimization levels substantially affect performance; specifically, standard models typically exhibit bias towards CISC (x86/x64) architectures and degrade under aggressive optimization, whereas reasoning models mitigate these biases and effectively refactor convoluted logic into cleaner code (e.g., achieving 72.31\% semantic preservation on ARM and reducing Halstead complexity to 27.08$\times10^4$ under O3, compared to 31.23$\times10^4$ for standard models). Fourth, In-context learning exhibits a counterintuitive divergence. Specifically, it significantly boosts performance for standard models (e.g., improving CodeLlama’s semantic preservation to 72.92\% in the 5-shot setting), yet proves counterproductive for reasoning models, where performance plateaus or degrades (e.g., DeepSeek-R1 saturates at 70.29\%) by interfering with their internal reasoning steps. Finally, our results show that LLMs’ deobfuscation capabilities can generalize effectively to more complex malicious binary scenarios, where reasoning models reduce code complexity by nearly 60\% and task-specific fine-tuning achieves the highest semantic preservation while neutralizing manually crafted obfuscation.

\noindent\textbf{Contributions.} We make the following contributions:
\begin{itemize}[leftmargin=2em]
    \setlength{\itemsep}{0pt}  
    \item To the best of our knowledge, this is the first systematic evaluation of LLMs for binary code deobfuscation. We construct a large-scale, rigorously quality-controlled dataset of C/C++ programs, leveraging diverse obfuscation transformations from mainstream open-source tools to enable a controlled and comprehensive assessment of current LLM capabilities.
    \vspace{0.3ex}
    \item We introduce \sysname, a comprehensive benchmark that evaluates deobfuscation results across four critical dimensions, i.e., lexical consistency, semantic preservation, code simplicity, and code readability, providing a principled measure of both deobfuscation effectiveness and practical utility.
    \vspace{0.3ex}
    \item We conduct extensive experiments on nine advanced LLMs, demonstrating their practical capabilities in binary code deobfuscation and deriving key empirical insights to guide the future design of LLM-based deobfuscation approaches.
\end{itemize}

\section{Background and Related Works}

In this section, we establish the context for our study. We first formalize the problem definition in $\S$\ref{problem definition}, followed by a review of related works in $\S$\ref{related works}.

\vspace{-1.0ex}
\subsection{Problem Definition} 
\label{problem definition}

Given source code $S$, $\mathcal{C}_{S \to I}$ transforms it to Intermediate Representation (IR) $I$, and $\mathcal{C}_{I \to B}$ generates the binary code $B$. We model the obfuscation process as a sequential pipeline where transformations $\mathcal{T}$ can be applied at the source, IR, or binary stages:
$$
S' = O_S(S, \mathcal{T}_S); \quad I' = O_I(\mathcal{C}_{S \to I}(S'), \mathcal{T}_I); \quad B' = O_B(\mathcal{C}_{I \to B}(I'), \mathcal{T}_B)
$$
where $S'$, $I'$, and $B'$ denote the resulting code states at each stage. When setting $\mathcal{T} = \emptyset$, the corresponding obfuscation function $O$ reduces to the identity mapping (i.e., $O(x, \emptyset)=x$), leaving that stage unobfuscated.
Regardless of the injection stage, the final output is an executable binary $B_{obf} = B'$ that is semantically equivalent to the original program $B$. In this work, the deobfuscator $D$ operates on the obfuscated pseudocode $P'$ derived from $B_{obf}$ to produce a recovered pseudocode $P'' = D(P')$. The objective is to ensure that $P''$ is semantically equivalent to the original unobfuscated pseudocode $P$ while restoring the clarity and readability of the code.

Note that in this paper, we focus on decompiled pseudocode rather than disassembled assembly code. This choice reflects practical reverse engineering workflows, especially in the analysis of maliciously obfuscated binaries, where analysts primarily rely on the pseudocode view produced by decompilers for a clearer and more structured representation. Modern decompilers automatically eliminate many low-level and redundant operations and generate pseudocode at a higher level of abstraction, making program semantics easier to understand by explicitly exposing control-flow structures, function boundaries, and variables recovered from low-level registers and memory locations. Consequently, pseudocode serves as a more suitable input for LLM-based deobfuscation.

\subsection{Related Works} \label{related works}

Existing deobfuscation approaches can be broadly categorized into three paradigms: static analysis, dynamic analysis, and learning-based techniques~\cite{shirazi2019analysis}. Static analysis~\cite{tofighi2019defeating, li2024x, tofighi2018dose, lee2024poster} focuses on inspecting code structure and control flow in the absence of execution. These approaches typically employ methods such as pattern matching, symbolic execution, and algebraic simplification to identify and eliminate obfuscation artifacts. For instance, Tofighi-Shirazi et al.~\cite{tofighi2018dose} propose DoSE, which statically analyzes syntax and semantics to identify functionally equivalent code fragments and merges redundant paths to simplify control flow. Conversely, dynamic analysis entails executing programs within a controlled environment to observe runtime behavior, thereby facilitating the recovery of semantic information obscured by static transformations. Many studies~\cite{menguy2021search, zhao2021input, david2020qsynth, blazytko2017syntia, tang2017seead} on binary code deobfuscation have concentrated on dynamic analysis. For example, Robin et al.~\cite{david2020qsynth} propose QSynth, a program synthesis framework that integrates dynamic symbolic execution with offline enumerative synthesis to simplify obfuscated expressions. 

Beyond these traditional analysis-based methods, recent studies have increasingly leveraged learning-based techniques, with LLMs showing promising performance in code deobfuscation~\cite{beste2025exploring, lachaux2021dobf, choi2024chatdeob, nataliealfredo}; however, prior work predominantly focused on the source code. For example, Beste et al.~\cite{beste2025exploring} demonstrate that fine-tuned LLMs can effectively reconstruct source code following composite obfuscation transformations and significantly reduce code complexity. In contrast, binary code deobfuscation poses fundamentally different challenges. Unlike source code, binaries undergo compilation and obfuscation steps that remove or obscure high-level semantic information, thereby making program structure and intent significantly harder to recover. The ability of LLMs to infer binary program logic and reconstruct deobfuscated pseudocode has not yet been systematically and quantitatively evaluated in this context. Recent studies~\cite{mohseni2025can, tkachenko2025deconstructing, tan2024llm4decompile} have begun to explore the potential of LLMs within the domain of binary obfuscation and deobfuscation. For instance, Mohseni et al.~\cite{mohseni2025can} demonstrated the generative capacity of LLMs for obfuscated assembly code using the METAMORPHASM framework, while Tkachenko et al.~\cite{tkachenko2025deconstructing} introduced a multi-dimensional framework to assess deobfuscation performance. Notably, Choi et al.~\cite{choi2024chatdeob} propose ChatDEOB, which utilizes a fine-tuned LLM to accurately reconstruct code affected by transformations such as Mixed Boolean-Arithmetic and Control Flow Flattening. While these efforts offer early evidence of LLM potential at the binary level, they remain limited in scope and do not provide a rigorous evaluation across diverse obfuscation settings, which necessitates our work.

\section{Overview}

In this section, we first describe the primary challenges in evaluating LLM-based binary code deobfuscation ($\S$\ref{challenges}). Building upon this, we articulate our insights and proposed solutions in $\S$\ref{insights and solutions}. Finally, we formalize these into the \sysname workflow, detailed in $\S$\ref{BinDeObfBench Workflow}.

\subsection{Challenges}
\label{challenges}

\noindent\textbf{C1: Lack of Realistic and Diverse Binary Obfuscation Dataset.} Existing evaluation efforts~\cite{mariano2024control, greco2024enabling, tkachenko2025deconstructing, feng2025debra, sebastian2017predicting, guthaus2001mibench, henning2006spec} frequently rely on limited datasets that predominantly focus on a single architecture or individual obfuscation transformations in isolation. In contrast, real-world software typically employs composite obfuscation strategies, where multiple transformations are applied in combination across different compilation stages. Consequently, current datasets fail to capture the complexity and heterogeneity found of practical obfuscation settings. This discrepancy hinders the comprehensive assessment of LLM robustness against complex obfuscation, thereby undermining the validation of their practical reliability.

\noindent\textbf{C2: Absence of Comprehensive Metrics for Deobfuscation Evaluation.} Existing research~\cite{tkachenko2025deconstructing, tofighi2018dose} on binary code deobfuscation lacks comprehensive metrics to effectively assess model performance. 
While effective deobfuscation aims to preserve program semantics while restoring code simplicity and readability, current evaluations predominantly emphasize semantic correctness alone. This narrow focus prevents a holistic assessment of deobfuscation quality and obscures trade-offs between semantic preservation and code readability in practice.

\noindent\textbf{C3: Knowledge Gap Caused by Distribution Shift.} Despite the remarkable success of LLMs in natural language and code generation, a significant knowledge gap persists regarding decompiled pseudocode derived from obfuscated binaries~\cite{tan2024llm4decompile, jin2023binary, shang2025empirical}. Obfuscation transformations introduce irregular control flows and high-entropy patterns that significantly deviate from the well-formed source code distributions encountered during LLM pre-training. This distribution shift hinders the models' ability to infer program logic and semantics from decompiled pseudocode, limiting their effectiveness in reverse engineering tasks.

\subsection{Insights and Solutions}
\label{insights and solutions}

\noindent\textbf{S1: A Large-scale and Diverse Obfuscation Evaluation Dataset.} To address the limitations of existing benchmarks, we construct from scratch a dataset of obfuscated binary code encompassing diverse instruction set architectures, compilation settings, and obfuscation transformations. These transformations are carefully selected to reflect techniques commonly adopted in real-world software protection, and are applied at three key stages of the software development pipeline: source code, intermediate representation, and binary. This multi-stage coverage enables the dataset to capture both structural and semantic complexities introduced by diverse obfuscation workflows.

\noindent\textbf{S2: Multi-dimensional Deobfuscation Assessment Metrics.} To address the absence of systematic evaluation metrics, we design a multi-dimensional assessment framework that evaluates deobfuscation performance from complementary perspectives, including lexical consistency, semantic preservation, code conciseness and readability. Specifically, we begin by evaluating the lexical consistency of the deobfuscated pseudocode using BLEU~\cite{papineni2002bleu}, ensuring that identifiers, keywords, and other lexical elements remain consistent with the unobfuscated pseudocode. Semantic preservation is subsequently quantified via a Dual-Perspective Semantic Fusion method, which integrates implicit semantic embeddings with explicit symbolic signatures to verify behavioral consistency. Code conciseness is measured using token-wise delta entropy, which measures the reduction of token-level uncertainty introduced by obfuscation. Finally, code readability is assessed using Halstead Complexity, which estimates the cognitive effort required to comprehend the deobfuscated pseudocode, thereby serving as an indirect indicator of its comprehensibility.

\noindent\textbf{S3: Bridging the Knowledge Gap of LLMs in Obfuscated Binary Understanding.} Existing studies~\cite{shang2024far, shang2025empirical} have shown that, even with only prompt engineering, LLMs can outperform expert-designed small models on specific binary understanding tasks. Building on these findings, we adopt in-context learning to mitigate the knowledge gap induced by distribution shift in binary code deobfuscation. Specifically, we provide LLMs with pairs of obfuscated code and their corresponding deobfuscated counterparts as in-context examples, enabling models to better adapt to obfuscated binary representations. Furthermore, we introduce Recopilot~\cite{chen2025recopilot} as a domain-specific expert LLM and ChatDEOB~\cite{choi2024chatdeob} as a task-specific expert LLM. Specifically, Recopilot is pre-trained across a broad spectrum of binary code analysis tasks excluding deobfuscation, whereas ChatDEOB is fine-tuned explicitly for the deobfuscation task. We evaluate their performance to dissect the distinct contributions of domain-specific versus task-specific expertise.

\begin{figure}[t]
  \centering
  \includegraphics[width=0.90\linewidth]{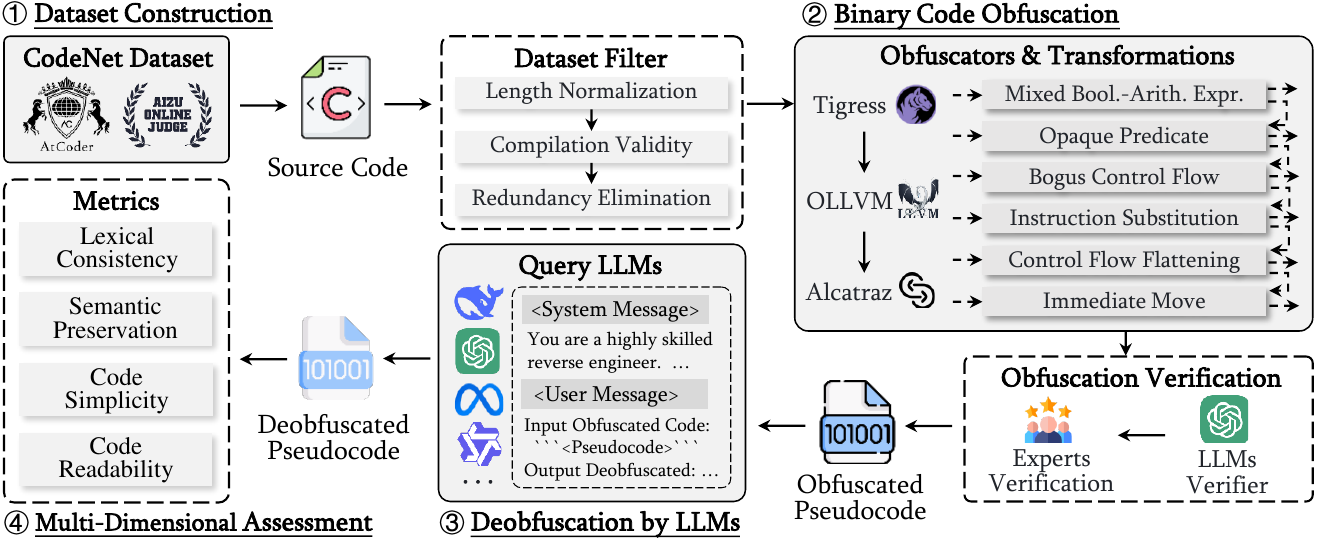}
  \caption{Workflow of \sysname}
  \vspace{-3.0ex}
  \label{fig:workflow}
\end{figure}

\subsection{{\bfseries\scshape BinDeObfBench} Workflow} \label{BinDeObfBench Workflow}

\autoref{fig:workflow} illustrates the pipeline of \sysname, comprising four key phases: (i) source code filtering and collection, (ii) generation of a high-quality obfuscated dataset through verified transformations, (iii) execution of the binary deobfuscation task using target LLMs, and (iv) systematic evaluation of performance across four metrics.

\section{Implementation Design}
\subsection{Dataset Construction} \label{dataset construction}

As outlined in Section \ref{challenges}, recognizing the limitations of the existing datasets, we design a set of filtering rules and data processing procedures to construct a obfuscated dataset from scratch.

\textit{Data Source Selection and Granularity.} To accurately model real-world software protection practices, we analyze the generation pipeline from source code to executable binaries, including pre-compilation, compile-time, and post-compilation stages. Different obfuscation transformations can be introduced at each phase, enabling a multi-layered representation of obfuscation strategies. Regarding data granularity, we select file-level source code as the obfuscated unit and prioritize it over repository-level projects. This decision is primarily driven by the inherent complexity of repository-level structures, including cross-file dependencies and intricate directory hierarchies. Directly applying obfuscation to source code at the repository-level introduces substantial technical challenges, high resource consumption, and intricate dependency conflicts, often resulting in compilation errors that are difficult to control. In contrast, file-level obfuscation offers better semantic validity and controllability, while remaining amenable to large-scale processing. Crucially, this granularity allows flexible application of different obfuscation transformations at the three stages mentioned above, thereby facilitating experimental operability and ensuring reproducibility.

For the data source, we utilize the CodeNet~\cite{puri2021codenet} dataset, which is a professionally curated and widely recognized benchmark, making it a reliable choice for our study. This choice also follows established in LLM evaluation, such as HumanEval~\cite{brown2020language} and EvalPlus~\cite{evalplus}, which similarly leverage programming challenge datasets as benchmarks. From CodeNet, we extract a subset of 754,058 C/C++ programs across 3,366 problems. This collection encompasses a broad spectrum of algorithmic types, capturing the logical complexity and diversity required to simulate real-world software scenarios for rigorous evaluation.

\textit{Dataset filtering.} While CodeNet provides an extensive collection of C/C++ programs, raw data contains noise that can compromise evaluation validity. Specifically, we identify three primary issues: (i) extreme length variations (excessively short or long samples) that bias LLM performance; (ii) compilation failures, which prevent binary generation; and (iii) data redundancy arising from multiple submissions for identical problems. To mitigate these issues, we employ a three-step filtering pipeline, distilling the initial dataset into a final set of 2,092 high-quality programs.

\begin{itemize}[leftmargin=1em]
    \item \textbf{Token Length Constraint.} We retain source files with lengths between 256 and 8,000 tokens, preserving approximately 95\% of the data. This range aligns with the length of typical real-world code~\cite{lu2021codexglue, husain1909codesearchnet} while remaining compatible with the context window limits of LLMs.
    \item \textbf{Compilation Validity Check.} We retain only samples that compile successfully, ensuring that each program can reliably pass through the subsequent obfuscation and decompilation stages.
    \item \textbf{Redundancy Elimination.} To maximize dataset diversity and minimize evaluation overhead, we deduplicate the dataset by selecting a single representative solution for each distinct problem.
\end{itemize}

\textit{Binary Code Obfuscation.} We collect four commonly used obfuscators: OLLVM~\cite{ollvm}, Hikari~\cite{Hikari}, Tigress~\cite{tigress}, and Alcatraz~\cite{Alcatraz}, representing both open-source and commercial obfuscation tools. \autoref{tab:obfuscators} summarizes the obfuscation transformations supported by them, along with their corresponding application phases. These tools are extensively adopted in academic research and industryial practice, covering all phases of obfuscation from source code to binary. Guided by prior studies~\cite{tkachenko2025deconstructing, mohseni2025can, david2020qsynth, tofighi2019defeating, mariano2024control, ming2015loop}, we restrict our scope to the following six mainstream obfuscation transformations. These methods span both form-based (expression-level) and structural (control-flow–level) obfuscation techniques commonly used in real-world software protection.

\begin{itemize}[leftmargin=1em]
    \item \textbf{Bogus Control Flow (BCF).} Inserts additional conditional branches, loops, or jump statements into the code that do not affect program semantics, creating spurious execution paths.
    \item \textbf{Instruction Substitution (SUB).} Replaces instructions with semantically equivalent alternatives, including arithmetic, logical, or data movement operations.
    \item \textbf{Control Flow Flattening (FLA).} Transforms the code's control flow into a flattened structure, typically using a central dispatcher or loop to manage all original branches.
    \item \textbf{Mixed Boolean-Arithmetic Expression (MBA).} Combines Boolean logic and arithmetic operations to produce more complex constructs while preserving original computations.
    \item \textbf{Opaque Predicate (Opaque).} Embeds conditional statements with predicates that always evaluate to a constant, creating misleading branches that are interleaved with normal code.
    \item \textbf{Immediate Move (ImmMov).} Decomposes a direct assignment or simple operation into multiple steps, producing a redundant expression that is semantically equivalent to the original operation.
\end{itemize}

\begin{wraptable}{hr}{8cm}
    \centering
    \caption{Obfuscators and Obfuscation Transformations.}
    \setlength{\tabcolsep}{0.99mm}
    \scalebox{0.75}{
    \begin{tabular}{lll}
        \toprule
            \textbf{Obfuscators} & \textbf{Transformations} & \textbf{Phase} \\
        \midrule
            OLLVM & \makecell[l]{Bogus Control Flow, \\ Instruction Substitution, \\ Control Flow Flattening} & \multirow{3}{*}{\centering \makecell[l]{Intermediate \\ Representation}} \\
        \cmidrule(r){1-2}
            Hikari & \makecell[l]{Anti-Class Dump, String Encryption, \\ Indirect Branching} & \\
        \midrule
            Tigress & \makecell[l]{Virtualize, Opaque Predicate, \\ Encode Branches, JitDynamic \\ Mixed Boolean-Arithmetic Expression} & Source Code \\ 
        \midrule
            Alcatraz & \makecell[l]{Obfuscation of Immediate Moves, \\ ADD Mutation, Lea Obfuscation} & Binary\\
        \bottomrule
    \end{tabular}}
    \label{tab:obfuscators} 
\end{wraptable}

\textit{Transformation Combinations.} To faithfully emulate the complexity of real-world software protection, we employ composite obfuscation transformations rather than relying solely on isolated techniques. Specifically, we organize the six selected transformations into varying complexity levels based on their combinations, ranging from single transformations (Level-1) to the simultaneous application of all six (Level-6). This combinatorial approach yields 63 distinct configurations (${\textstyle \sum_{k=1}^{6} C_{6}^{k}}$), which, when applied to the 2,092 filtered samples, generate 131,796 obfuscated bianries. Furthermore, we also consider four ISAs (ARM, MIPS, x86, x64) and four optimization options (O0-O3), applying each obfuscation combination across all architectures and optimization settings. This process generates a total of 1,564,816 stripped binaries, where symbol and debug information are removed to better reflect realistic reverse engineering scenarios.

\textit{Validity Verification and Sampling.} Ensuring the effective application of obfuscation is a prerequisite for reliable evaluation. Given that transformations are not always successfully applied, we implement a two-stage verification pipeline to guarantee data quality. First, we perform string matching to filter out ineffective transformations by discarding samples where the obfuscated pseudocode remains lexically identical to the original. Second, we utilize GPT-4o as an automated verifier to assess whether the specific obfuscation type is correctly applied by presenting the model with triplets of <\textit{original pseudocode}, \textit{obfuscation type}, \textit{obfuscated pseudocode}>. From the resulting pool of verified candidates, we employ a stratified random sampling strategy to select 500 distinct pseudocode functions for each unique configuration of ISA, optimization option, and obfuscation type, forming the benchmark for our evaluation. This approach balances computational efficiency with statistical significance while ensuring sample uniqueness.

\begin{wraptable}{hr}{8.0cm}
    \centering
    \caption{The Results of Pilot Study.}
    \setlength{\tabcolsep}{1.2mm}
    \scalebox{0.98}{
    \begin{threeparttable}
        \renewcommand{\arraystretch}{1.0}
        \begin{tabular}{lcc}
            \toprule
                 & \textbf{Obfuscated$_{G}$\tnote{1}} & \textbf{Unobfuscated$_{G}$} \\
            \midrule
                \textbf{Obfuscated$_{H}$\tnote{2}} & 468 & 21 \\
                \textbf{Unobfuscated$_{H}$} & 3 & 8 \\
            \bottomrule
        \end{tabular}
        \begin{tablenotes}
            \footnotesize
            \item[1] $G$ denotes GPT-4o, the verifier.
            \item[2] $H$ denotes the ground truth provided by reverse engineers.
        \end{tablenotes}
    \end{threeparttable}}
    \label{tab:preliminary experiment} 
\end{wraptable}

To validate the reliability of this automated verifier, we conducted a pilot study by randomly selecting 500 pseudocode functions from verified candidates. In this pilot, three reverse engineering experts independently assessed the samples, yielding a Fleiss' kappa~\cite{fleiss1971measuring} of 0.83, indicating almost perfect agreement. Any discrepancies were resolved through consensus to establish the ground truth. As shown in \autoref{tab:preliminary experiment}, GPT-4o achieved an overall accuracy of 95.2\%. Notably, error analysis reveals a conservative bias: the model occasionally rejects valid obfuscated samples but rarely misclassifies unobfuscated code as valid. This behavior is desirable for dataset construction, as it prioritizes precision over recall and minimizes false positives. Overall, these findings confirm the high reliability of GPT-4o as a gatekeeper for dataset construction.

\textit{Malware Dataset.} To evaluate LLM deobfuscation capabilities within realistic threat scenarios, we also collect samples from several representative malware families, including Backdoors, Botnets, and Trojans, from two well-known open-source repositories, theZoo~\cite{theZoo} and MalwareSourceCode~\cite{MalwareSourceCode}. To better approximate realistic attack conditions, these samples are recompiled with obfuscation and stripped of all symbolic information. Subsequently, we apply the aforementioned filtering criteria and validity verification pipeline to maintain data quality. Ultimately, we select a random subset of 500 validated functions to constitute the malware evaluation dataset.

\begin{table}[t]
  \centering
  \renewcommand{\arraystretch}{1.1}
  \caption{The Detailed Information of Large Language Models Employed in this Research.}
  \setlength{\tabcolsep}{0.8mm}{
  \scalebox{0.80}{
  \begin{threeparttable}
    \begin{tabular}{llclclcc}
    \toprule
        \textbf{Domain} & \multicolumn{1}{l}{\textbf{Model}} & \textbf{Size} & \multicolumn{1}{c}{\textbf{Model ID}} & \textbf{\makecell{Context \\ Length}} & \textbf{\makecell{Base Model}} & \textbf{Publisher} & \textbf{License\tnote{1}} \\
    \midrule
        \multirow{2}{*}{Code-Specific} & Qwen2.5-Coder~\cite{hui2024qwen2} & 32B & Qwen2.5-Coder-32B-Instruct & 128K & Qwen2.5-Coder-32B & Qwen & $\bullet$ \\
        & CodeLlama~\cite{codellama} & 70B & CodeLlama-70b-Instruct-hf & 16K & CodeLlama-70b-hf & Meta AI& $\bullet$ \\
        \cmidrule{1-8}
        \multirow{4}{*}{General-Purpose} & Qwen3~\cite{qwen3technicalreport} & 32B & Qwen3-32B & 32K & Qwen3-32B-Base & Qwen & $\bullet$ \\
        & Llama3.1~\cite{llama3.1} & 70B & Llama-3.1-70B-Instruct & 128K & Llama-3.1-70B & Meta AI & $\bullet$ \\
        & DeepSeek-V3~\cite{liu2024deepseek} & 671B & DeepSeek-V3-0324 & 64K & DeepSeek-V3-Base & DeepSeek & $\bullet$ \\
        & GPT-4o~\cite{gpt-4o} & - & GPT-4o-2024-11-20 & 128K & GPT-4 & OpenAI & $\circ$ \\
        \cmidrule{1-8}
        \multirow{2}{*}{Reasoning-Optimized} & DeepSeek-R1~\cite{guo2025deepseek} & 671B & DeepSeek-R1-0528 & 64K & DeepSeek-V3-Base & DeepSeek & $\bullet$ \\
        & OpenAI-o1~\cite{opeai-o1} & - & OpenAI-o1 & 128K & - & OpenAI & $\circ$ \\
        \cmidrule{1-8}
        Domain-Specific Expert\tnote{2} & Recopilot~\cite{chen2025recopilot} & 7B & recopilot-v0.1-beta-dpo & 128K & Qwen2.5-Coder-7B & QIANXIN & $\circ$ \\
        \cmidrule{1-8}
        Task-Specific Expert\tnote{2} & ChatDEOB~\cite{choi2024chatdeob} & 7B & Qwen2.5-Coder-7B-Instruct & 128K & Qwen2.5-Coder-7B & - & $\circ$ \\
    \bottomrule
    \end{tabular}
    \begin{tablenotes}
        \footnotesize
        \item[1] "$\bullet$" indicates Open-Source, "$\circ$" indicates Closed-Source. \hspace{2ex} $^2$ Domain- and Task-Specific Experts refer to LLMs pre-trained on general binary analysis (excluding deobfuscation) and those explicitly fine-tuned for deobfuscation, respectively.
    \end{tablenotes}
    \end{threeparttable}}}
    \label{tab:llmsinfo} 
    \vspace{-3.0ex}
\end{table}

\subsection{Deobfuscation by LLMs}

To evaluate the capability of LLMs in deobfuscating binary code, we design task-oriented prompts that leverage their in-context learning~\cite{zheng2023survey}.

\textit{Prompt Engineering.} We employ two distinct prompting strategies, i.e., zero-shot and few-shot, to assess model performance under varying levels of guidance. In the zero-shot setting, we adhere to standard practices by providing only the task description and the target obfuscated pseudocode. However, given the significant distribution shift between obfuscated binary pseudocode and general source code, as noted in Challenge C3, zero-shot inference may struggle to capture complex obfuscation patterns. To bridge this gap, we implement a few-shot strategy that augments the prompt with demonstration examples consisting of aligned pairs of obfuscated and deobfuscated code. Prior research~\cite{shang2024far, brown2020language} has shown that such in-context demonstrations are particularly effective for adapting LLMs to specialized domains like reverse engineering. Relative to more complex paradigms (e.g., self-consistency~\cite{wang2022self}), the few-shot strategy achieves comparable performance while incurring significantly lower computational overhead.

\textit{LLMs Selection.} To ensure a comprehensive evaluation, we categorize LLMs into five distinct groups: general-purpose, code-specific, reasoning-optimized, domain-specific expert, and task-specific LLMs. From each of these categories, we select representative models to form our evaluation testbed, with their detailed specifications summarized in \autoref{tab:llmsinfo}. Notably, our implementation of the ChatDEOB~\cite{choi2024chatdeob} differs from the original paper, which uses GPT-3.5-Turbo~\cite{gpt-3.5-turbo} as the backbone. We instead adopt Qwen2.5-Coder-7B-Instruct~\cite{hui2024qwen2}, based on two main considerations. First, we prioritize open-source models to avoid the high costs and accessibility limitations associated with fine-tuning closed-source models. Second, to fairly compare the impact of domain-specific training versus task-specific fine-tuning, we align the backbone with that of Recopilot. In addition, we strictly apply their methodology on a dataset of equivalent scale constructed from our corpus and fine-tune the LLM accordingly, ensuring that the training data is distinct from our benchmark.

\textit{Non-LLM Deobfuscation Methods.} To offer a comprehensive perspective on deobfuscation performance, we incorporate existing non-LLM methods for comparative analysis. While recent studies~\cite{choi2024chatdeob, li2024x, menguy2021search, david2020qsynth, tofighi2019defeating, tofighi2018dose, zhao2021input, tang2017seead, blazytko2017syntia, lee2024poster} have made significant strides in this domain, the majority of these approaches remain closed-source, which precludes direct and reproducible comparison (as detailed in \autoref{tab:non-llm}). Among the few open-source methods, Xyntia~\cite{menguy2021search} and QSynthesis~\cite{david2020qsynth} are not suitable due to inherent limitations. Specifically, Xyntia is limited to processing code snippets and cannot handle complete functions or binary programs, while QSynthesis necessitates the manual identification of obfuscated regions, making it impractical for benchmarking at scale. For practical and scalable evaluation, we therefore focus on D810~\cite{D810} and GooMBA~\cite{GooMBA}, which are widely adopted in real-world reverse engineering tasks. Functionally, D810 hooks into the Hex-Rays microcode pipeline, combining backward variable tracing, simulated execution, and pattern matching within the intermediate representation layer to restructure obfuscated code and recover semantics. GooMBA leverages expression tree traversal, algebraic simplification, heuristic evaluation, and SMT solver verification within the same layer to automatically identify and simplify Mixed Boolean-Arithmetic expressions with guaranteed semantic equivalence.

\begin{table}[t]
    \centering
    \caption{Summary of Existing Binary Code Deobfuscation Methods.}
    \renewcommand{\arraystretch}{0.95}
    {\renewcommand\cellset{\renewcommand\arraystretch{0.7}}}
    \setlength{\tabcolsep}{0.98mm}
    \scalebox{0.80}{
    \begin{threeparttable}
    \begin{tabular}{ccccc||ccccc}
        \toprule
        \textbf{Method} & \textbf{Object} & \textbf{Paradigm} & \textbf{Year} & \textbf{License} & \textbf{Method} & \textbf{Object} & \textbf{Paradigm} & \textbf{Year} & \textbf{License}\tnote{1} \\
        \midrule
        ChatDEOB~\cite{choi2024chatdeob} & \makecell{MBA, FLA, \\Opaque} & \makecell{Learning-\\based} & 2024 & $\circ$ & AutoSimpler~\cite{zhao2021input} & \makecell{SUB, EncodeArithmetic} & Dynamic & 2021 & $\circ$ \\
        EMBA~\cite{lee2024poster} & \makecell{MBA} & Static & 2024 & $\circ$ & QSynthesis~\cite{david2020qsynth} & \makecell{MBA, Virtualizaion, \\Data Encoding} & Dynamic & 2020 & $\bullet$\\
        X-MBA~\cite{li2024x} & MBA & Static & 2024 & $\circ$ & Deobfuscator~\cite{tofighi2019defeating}\tnote{2} & Opaque & Static & 2019 & $\circ$ \\
        gooMBA~\cite{GooMBA} & MBA & Static & 2023 & $\bullet$ & DoSE~\cite{tofighi2018dose} & \makecell{Opaque, Code Clone, \\Range Dividers} & Static & 2018 & $\circ$ \\
        Xyntia~\cite{menguy2021search} & \makecell{MBA, Virtualization, \\Opaque, Path Explosion} & Dynamic & 2021 & $\bullet$ & SEEAD~\cite{tang2017seead} & \makecell{FLA, Virtualization} & Dynamic & 2017 & $\circ$ \\
        D810~\cite{D810} & \makecell{BCF, FLA, SUB} & Static & 2021 & $\bullet$ & Syntia~\cite{blazytko2017syntia} & \makecell{MBA, Virtualization} & Dynamic & 2017 & $\circ$ \\
        \bottomrule 
    \end{tabular}
    \begin{tablenotes}
        \footnotesize
        \item[1] "$\bullet$" indicates Open-Source, "$\circ$" indicates Closed-Source. \hspace{2ex}  $^2$ We denote the unnamed method as "Deobfuscator".
    \end{tablenotes}
    \end{threeparttable}
    }
    \vspace{-3.0ex}
    \label{tab:non-llm}
\end{table}

\subsection{Multi-Dimensional Assessment Framework}

Previous studies~\cite{tkachenko2025deconstructing} have relied on limited metrics, making it challenging to perform a comprehensive assessment of deobfuscation efficacy. To address this, we establish a multi-dimensional evaluation framework that assesses code quality across four complementary dimensions: \textit{lexical consistency}, \textit{semantic preservation}, \textit{code simplicity}, and \textit{code readability}. Each dimension targets a specific property of deobfuscated code, allowing us to evaluate both semantic correctness and practical code quality in a unified framework. The four dimensions are described as follows:

\begin{wrapfigure}{r}{0.48\linewidth}
  \centering
  \scalebox{1.0}{
    \includegraphics[width=\linewidth]{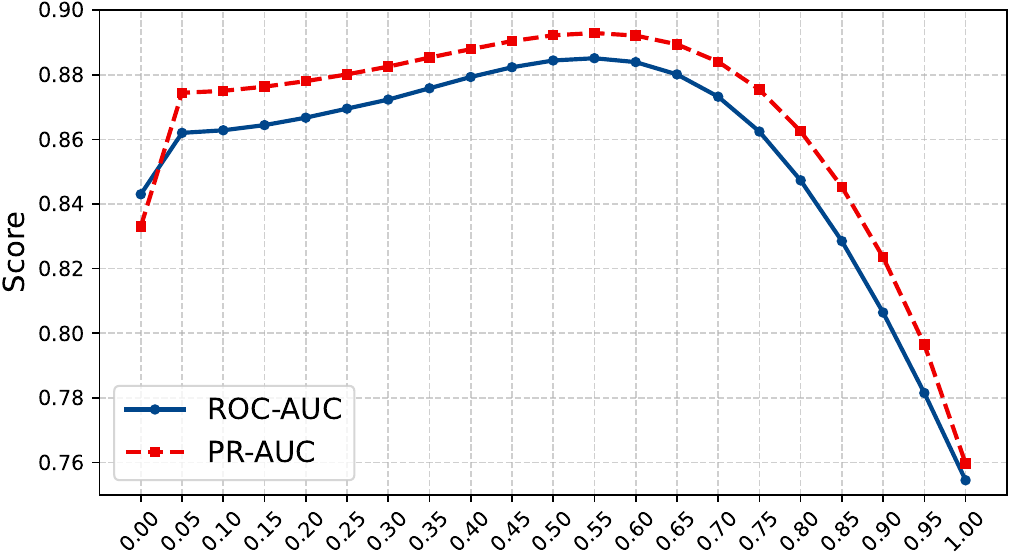}
  }
  \caption{Grid search for optimal $\alpha$.}
  \label{fig:alpha_auc}
  \vspace{-3.0ex}
\end{wrapfigure}

\textit{(1) Lexical Consistency.} Since binary code deobfuscation aims to recover a readable representation that closely reflects the original program, lexical consistency provides a direct measure of how faithfully the recovered code matches the unobfuscated pseudocode. Specifically, it evaluates alignment at the textual level, including identifier naming, function signatures, and syntactic structure. We quantify lexical consistency using the BLEU~\cite{papineni2002bleu} score, which measures n-gram overlap between the deobfuscated output and the reference pseudocode.

\textit{(2) Semantic Preservation.} Assessing semantic consistency before and after deobfuscation is challenging because decompiled pseudocode is non-executable, rendering traditional dynamic verification techniques based on test execution inapplicable. To address this, we propose a Dual-Perspective Semantic Fusion method, detailed in Algorithm \ref{algorithm1}, which conceptualizes function semantics as a synthesis of implicit and explicit features. To capture implicit semantics, inspired by prior research utilizing LLMs as semantic encoders~\cite{behnamghader2024llm2vec, nie2024text, liu2024codexembed, kryvosheieva2025efficient}, we adopt Qwen2.5-Coder-1.5B-Instruct~\cite{hui2024qwen2, qwen2} as the backbone, which is further fine-tuned via domain-adaptive pre-training on 112,000 pseudocode and contrastive learning on 62,400 <\textit{obfuscated}, \textit{unobfuscated}> pairs. The resulting embedding similarity reflects implicit semantic correspondence beyond surface syntax.

\begin{wrapfigure}{r}{0.565\textwidth}
\scalebox{0.95}{
    \begin{algorithm}[H]
    \label{algorithm1}
    \caption{Dual-Perspective Semantic Fusion}
    \footnotesize
    \SetKwInput{KwInput}{Input}
    \SetKwInput{KwParam}{Param}
    \SetKwInput{KwOutput}{Output}
    
    \KwInput{Original pseudocode $P_{gt}$, deobfuscated pseudocode $P_{deob}$}
    \KwParam{Encoder $\mathcal{M}$ (Qwen-Coder), fusion weight $\alpha$}
    \KwOutput{Semantic consistency score $S_{final}$}
    
    \tcp{Implicit semantic}
    $v_{gt}, v_{deob} \leftarrow \mathcal{M}(P_{gt}), \mathcal{M}(P_{deob})$\;
    $S_{emb} \leftarrow \mathrm{CosineSimilarity}(v_{gt}, v_{deob})$\;
    
    \tcp{Explicit semantic}
    $E_{gt}, E_{deob} \leftarrow \mathrm{ExtractEntities}(P_{gt}), \mathrm{ExtractEntities}(P_{deob})$\;
    $S_{jac} \leftarrow \dfrac{|E_{gt} \cap E_{deob}|}{|E_{gt} \cup E_{deob}|}$\;
    
    \tcp{Linear weighted fusion}
    $S_{final} \leftarrow \alpha \cdot S_{emb} + (1-\alpha) \cdot S_{jac}$\;
    
    \Return{$S_{final}$}
    \end{algorithm}
}
\vspace{-2.0ex}
\end{wrapfigure}

In parallel, we extract explicit semantic features that are relatively resilient to obfuscation, including input parameters, global variables, function calls, and return values against obfuscation transformations, calculating their Jaccard similarity~\cite{leskovec2020mining} to quantify such consistency. We integrate the implicit and explicit semantic scores through a linear fusion strategy, controlled by a weighting coefficient $\alpha$. Crucially, rather than enforcing a binary judgment of equivalence, this fused metric facilitates a comparative evaluation strategy to gauge the degree of semantic restoration. We calibrate the fusion coefficient $\alpha$, based on two complementary metrics: ROC-AUC, which assesses the global discriminative ability of the metric, and PR-AUC, which evaluates the robustness of positive identification under varying thresholds. We conduct a grid search on a validation set of 24,960 samples, with results summarized in \autoref{fig:alpha_auc}. As $\alpha$ varies, performance exhibits a clear inverted-U-shaped trend. Relying on a single semantic perspective (i.e., $\alpha = 0.00$ or $\alpha = 1.00$) yields suboptimal results, with ROC-AUC scores of 84.30\% and 75.45\%, respectively, whereas the fused metric achieves an optimal balance at $\alpha=0.55$ by reaching an ROC-AUC of 88.51\% and a PR-AUC of 89.29\%. Evaluation on an independent test set of the same scale further improves performance to ROC-AUC = 92.53\% and PR-AUC = 92.57\%, confirming the complementary nature of implicit and explicit semantic features. All data used for metric construction and calibration are drawn exclusively from the dataset described in $\S$\ref{dataset construction} and are strictly disjoint from \sysname.

\textit{(3) Code Simplicity.} To quantify the simplicity of deobfuscated pseudocode, we employ token-wise delta entropy, a variant of Shannon entropy inspired by prior work~\cite{mohseni2025can} that captures the incremental contribution of each token to the sequence's information complexity. We evaluate the metric across unobfuscated, obfuscated, and deobfuscated counterparts to measure the restoration of code simplicity. Formally, let $ P = [t_1, t_2, ..., t_n] $ denote a pseudocode token sequence, the token-wise delta entropy is defined as: 
$$\Delta H(t_i) = H([t_1,...,t_i]) - H([t_1,...,t_{i-1}]), i = 1,...,n$$
where $H(\cdot)$ represents the Shannon entropy of the prefix subsequence. The cumulative complexity of the sequence is given by $\Delta H(P) = \sum_{i=1}^{n} \Delta H(t_i)$. Consequently, a reduction in $\Delta H(P)$ after deobfuscation signifies the successful mitigation of obfuscation-induced complexity.

\textit{(4) Code Readability.} The primary objective of deobfuscation is to restore code readability and facilitate program comprehension. While human evaluation constitutes the gold standard for this task, it is prohibitively labor-intensive and impractical for large-scale analysis. To overcome this scalability limitation, we adopt Halstead complexity~\cite{munson1989dimensionality, hariprasad2017software} as an automated proxy. This metric quantifies the cognitive effort required to comprehend software based on the count and diversity of its operators and operands. Within this framework, \textit{Volume} measures the information content, \textit{Difficulty} reflects the cognitive complexity of understanding it, and \textit{Effort} estimates the mental exertion required for review and modification. In this work, we utilize \textit{Effort} as the primary indicator of readability. Unlike token-wise delta entropy, which gauges simplification via information density, Halstead complexity offers a more human-oriented perspective by quantifying the effort involved in reading and understanding the code’s lexical composition.

\vspace{-1.0ex}
\section{Evaluation}
\vspace{-1.0ex}

We design and conduct experiments to evaluate the performance of LLMs in decompiled pseudocode deobfuscation tasks and analyze them in relation to the following \textbf{Research Questions (RQs):}

\begin{itemize}[leftmargin=1em]
    \item \textbf{RQ1:} What is the performance of LLMs in binary code deobfuscation?
    \vspace{-0.5ex}
    \item \textbf{RQ2:} How do LLMs perform under different combinations of obfuscation transformations?
    \vspace{-0.5ex}
    \item \textbf{RQ3:} How do LLMs perform in deobfuscation across different architectures and optimizations?
    \vspace{-0.5ex}
    \item \textbf{RQ4:} What is the impact of in-context learning on the deobfuscation performance of LLMs?
    \vspace{-0.5ex}
    \item \textbf{RQ5:} How effective are LLMs at deobfuscating real-world malicious binaries?
\end{itemize}

\noindent\textbf{Platform.} All experiments are performed on a server running Ubuntu 24.10 with 256 CPU cores, 436 GB RAM, 100TB disk storage, and 8 NVIDIA RTX A6000 GPUs. The compilation, obfuscation, and decompilation of the \sysname dataset require approximately 15 days in total. In our evaluation, the Qwen and Llama model series are locally deployed, while other models are accessed via remote APIs. All reported results are averaged over three independent runs for each LLM.
\vspace{-0.5ex}

\begin{table}[t]
    \centering
    \caption{Overall Deobfuscation Performance of Evaluated Methods. \textcolor{customred}{\textbf{$\uparrow$}}/\textcolor{customgreen}{\textbf{$\downarrow$}} denote absolute increase and decrease.}
    \renewcommand{\arraystretch}{0.87}
    \setlength{\tabcolsep}{1.3mm}
    \scalebox{0.83}{
    \begin{threeparttable}
    \begin{tabular}{l|llll|llll}
        \toprule
        \makecell[l]{\textbf{Models}} & \makecell[c]{\textbf{Lexical${\textcolor{customred}{\textbf{$(\uparrow)$}}}$}\\\textbf{Consistency}} & \makecell[c]{\textbf{Semantic${\textcolor{customred}{\textbf{$(\uparrow)$}}}$}\\\textbf{Preservation}} & \makecell[c]{\textbf{Code${\textcolor{customgreen}{\textbf{$(\downarrow)$}}}$}\\\textbf{Simplicity}} & \makecell[c]{\textbf{Code}${\textcolor{customgreen}{\textbf{$(\downarrow)$}}}$\\\textbf{Readability}} & \makecell[c]{\textbf{Lexical${\textcolor{customred}{\textbf{$(\uparrow)$}}}$}\\\textbf{Consistency}} & \makecell[c]{\textbf{Semantic${\textcolor{customred}{\textbf{$(\uparrow)$}}}$}\\\textbf{Preservation}} & \makecell[c]{\textbf{Code${\textcolor{customgreen}{\textbf{$(\downarrow)$}}}$}\tnote{1}\\\textbf{Simplicity}} & \makecell[c]{\textbf{Code${\textcolor{customgreen}{\textbf{$(\downarrow)$}}}$}\tnote{2}\\\textbf{Readability}} \\
        \midrule
         & \multicolumn{4}{c|}{\textit{\textbf{Bogus Control Flow (BCF)}}} & \multicolumn{4}{c}{\textit{\textbf{Instruction Substitution (SUB)}}} \\
        \midrule
        Obf.Pseudocode & \multicolumn{1}{c}{\cellcolor{gray!30}33.24 (\%)} & \multicolumn{1}{c}{\cellcolor{gray!30}71.04 (\%)} & \multicolumn{1}{c}{\cellcolor{gray!30}5.70} & \multicolumn{1}{c|}{\cellcolor{gray!30}19.59 ($\times 10^4$)} & \multicolumn{1}{c}{\cellcolor{gray!30}87.03 (\%)} & \multicolumn{1}{c}{\cellcolor{gray!30}86.15 (\%)} & \multicolumn{1}{c}{\cellcolor{gray!30}5.62} & \multicolumn{1}{c}{\cellcolor{gray!30}9.56 ($\times 10^4$)} \\
        \midrule
        Qwen2.5-Coder & 49.38$_{\textcolor{customred}{\textbf{$(\uparrow16.14)$}}}$ & \underline{73.34}$_{\textcolor{customred}{\textbf{$(\uparrow2.30)$}}}$ & 5.63$_{\textcolor{customgreen}{\textbf{$(\downarrow0.07)$}}}$ & 11.60$_{\textcolor{customgreen}{\textbf{$(\downarrow7.99)$}}}$ & 85.26$_{\textcolor{customgreen}{\textbf{$(\downarrow1.77)$}}}$ & 86.27$_{\textcolor{customred}{\textbf{$(\uparrow0.12)$}}}$ & 5.57$_{\textcolor{customgreen}{\textbf{$(\downarrow0.05)$}}}$ & \hspace{1ex}8.49$_{\textcolor{customgreen}{\textbf{$(\downarrow1.07)$}}}$ \\
        Qwen3 & 37.59$_{\textcolor{customred}{\textbf{$(\uparrow4.35)$}}}$ & 72.11$_{\textcolor{customred}{\textbf{$(\uparrow1.07)$}}}$ & 5.68$_{\textcolor{customgreen}{\textbf{$(\downarrow0.02)$}}}$ & 17.92$_{\textcolor{customgreen}{\textbf{$(\downarrow1.67)$}}}$ & 87.90$_{\textcolor{customred}{\textbf{$(\uparrow0.87)$}}}$ & \underline{87.54}$_{\textcolor{customred}{\textbf{$(\uparrow1.39)$}}}$ & 5.61$_{\textcolor{customgreen}{\textbf{$(\downarrow0.01)$}}}$ & \hspace{1ex}9.37$_{\textcolor{customgreen}{\textbf{$(\downarrow0.19)$}}}$ \\
        Codellama & 44.46$_{\textcolor{customred}{\textbf{$(\uparrow11.22)$}}}$ & 71.52$_{\textcolor{customred}{\textbf{$(\uparrow0.48)$}}}$ & 5.41$_{\textcolor{customgreen}{\textbf{$(\downarrow0.29)$}}}$ & \hspace{1ex}8.90$_{\textcolor{customgreen}{\textbf{$(\downarrow10.69)$}}}$ & 67.39$_{\textcolor{customgreen}{\textbf{$(\downarrow19.64)$}}}$ & 83.91$_{\textcolor{customgreen}{\textbf{$(\downarrow2.24)$}}}$ & 5.42$_{\textcolor{customgreen}{\textbf{$(\downarrow0.20)$}}}$ & \hspace{1ex}6.79$_{\textcolor{customgreen}{\textbf{$(\downarrow2.77)$}}}$ \\
        Llama-3.1 & 25.26$_{\textcolor{customgreen}{\textbf{$(\downarrow7.98)$}}}$ & 61.90$_{\textcolor{customgreen}{\textbf{$(\downarrow9.14)$}}}$ & 5.42$_{\textcolor{customgreen}{\textbf{$(\downarrow0.28)$}}}$ & 14.57$_{\textcolor{customgreen}{\textbf{$(\downarrow5.02)$}}}$ & 62.51$_{\textcolor{customgreen}{\textbf{$(\downarrow24.52)$}}}$ & 76.05$_{\textcolor{customgreen}{\textbf{$(\downarrow10.10)$}}}$ & 5.35$_{\textcolor{customgreen}{\textbf{$(\downarrow0.27)$}}}$ & \hspace{1ex}7.02$_{\textcolor{customgreen}{\textbf{$(\downarrow2.54)$}}}$ \\
        DeepSeek-V3 & 44.89$_{\textcolor{customred}{\textbf{$(\uparrow11.65)$}}}$ & 71.30$_{\textcolor{customred}{\textbf{$(\uparrow0.26)$}}}$ & 5.38$_{\textcolor{customgreen}{\textbf{$(\downarrow0.32)$}}}$ & 12.16$_{\textcolor{customgreen}{\textbf{$(\downarrow7.43)$}}}$ & 65.39$_{\textcolor{customgreen}{\textbf{$(\downarrow21.64)$}}}$ & 79.84$_{\textcolor{customgreen}{\textbf{$(\downarrow6.31)$}}}$ & 5.38$_{\textcolor{customgreen}{\textbf{$(\downarrow0.24)$}}}$ & \hspace{1ex}7.30$_{\textcolor{customgreen}{\textbf{$(\downarrow2.26)$}}}$ \\
        GPT-4o & 36.41$_{\textcolor{customred}{\textbf{$(\uparrow3.17)$}}}$ & 68.08$_{\textcolor{customgreen}{\textbf{$(\downarrow2.96)$}}}$ & 5.52$_{\textcolor{customgreen}{\textbf{$(\downarrow0.18)$}}}$ & 11.97$_{\textcolor{customgreen}{\textbf{$(\downarrow7.62)$}}}$ & 66.68$_{\textcolor{customgreen}{\textbf{$(\downarrow20.35)$}}}$ & 79.03$_{\textcolor{customgreen}{\textbf{$(\downarrow7.12)$}}}$ & 5.63$_{\textcolor{customred}{\textbf{$(\uparrow0.01)$}}}$ & \hspace{1ex}9.26$_{\textcolor{customgreen}{\textbf{$(\downarrow0.30)$}}}$ \\
        DeepSeek-R1 & 51.15$_{\textcolor{customred}{\textbf{$(\uparrow17.91)$}}}$ & 69.38$_{\textcolor{customgreen}{\textbf{$(\downarrow1.66)$}}}$ & 5.33$_{\textcolor{customgreen}{\textbf{$(\downarrow0.37)$}}}$ & \hspace{1ex}\underline{7.20}$_{\textcolor{customgreen}{\textbf{$(\downarrow12.39)$}}}$ & 69.35$_{\textcolor{customgreen}{\textbf{$(\downarrow17.68)$}}}$ & 81.61$_{\textcolor{customgreen}{\textbf{$(\downarrow4.54)$}}}$ & 5.37$_{\textcolor{customgreen}{\textbf{$(\downarrow0.25)$}}}$ & \hspace{1ex}6.96$_{\textcolor{customgreen}{\textbf{$(\downarrow2.60)$}}}$ \\
        OpenAI-o1 & 35.98$_{\textcolor{customred}{\textbf{$(\uparrow2.74)$}}}$ & 70.27$_{\textcolor{customgreen}{\textbf{$(\downarrow0.77)$}}}$ & 5.36$_{\textcolor{customgreen}{\textbf{$(\downarrow0.34)$}}}$ & 10.12$_{\textcolor{customgreen}{\textbf{$(\downarrow9.47)$}}}$ & 62.27$_{\textcolor{customgreen}{\textbf{$(\downarrow24.76)$}}}$ & 80.11$_{\textcolor{customgreen}{\textbf{$(\downarrow6.04)$}}}$ & \underline{5.31}$_{\textcolor{customgreen}{\textbf{$(\downarrow0.31)$}}}$ & \hspace{1ex}\underline{6.45}$_{\textcolor{customgreen}{\textbf{$(\downarrow3.11)$}}}$ \\
        ReCopilot & 33.10$_{\textcolor{customgreen}{\textbf{$(\downarrow0.14)$}}}$ & 71.17$_{\textcolor{customred}{\textbf{$(\uparrow0.13)$}}}$ & 5.69$_{\textcolor{customgreen}{\textbf{$(\downarrow0.01)$}}}$ & 19.57$_{\textcolor{customgreen}{\textbf{$(\downarrow0.02)$}}}$ & 86.70$_{\textcolor{customgreen}{\textbf{$(\downarrow0.33)$}}}$ & 85.80$_{\textcolor{customgreen}{\textbf{$(\downarrow0.35)$}}}$ & 5.63$_{\textcolor{customred}{\textbf{$(\uparrow0.01)$}}}$ & \hspace{1ex}9.55$_{\textcolor{customgreen}{\textbf{$(\downarrow0.01)$}}}$ \\
        ChatDEOB & \underline{63.97}$_{\textcolor{customred}{\textbf{$(\uparrow30.73)$}}}$ & 72.45$_{\textcolor{customred}{\textbf{$(\uparrow1.41)$}}}$ & \underline{5.30}$_{\textcolor{customgreen}{\textbf{$(\downarrow0.40)$}}}$ & 11.49$_{\textcolor{customgreen}{\textbf{$(\downarrow8.10)$}}}$ & 89.09$_{\textcolor{customred}{\textbf{$(\uparrow2.06)$}}}$ & 82.63$_{\textcolor{customgreen}{\textbf{$(\downarrow3.52)$}}}$ & 5.56$_{\textcolor{customgreen}{\textbf{$(\downarrow0.06)$}}}$ & \hspace{1ex}9.57$_{\textcolor{customred}{\textbf{$(\uparrow0.01)$}}}$ \\
        \midrule
        D810 & 60.10$_{\textcolor{customred}{\textbf{$(\uparrow26.86)$}}}$ & 70.20$_{\textcolor{customgreen}{\textbf{$(\downarrow0.84)$}}}$ & 5.45$_{\textcolor{customgreen}{\textbf{$(\downarrow0.25)$}}}$ & \hspace{1ex}8.63$_{\textcolor{customgreen}{\textbf{$(\downarrow10.96)$}}}$ & \underline{93.01}$_{\textcolor{customred}{\textbf{$(\uparrow5.98)$}}}$ & 83.71$_{\textcolor{customgreen}{\textbf{$(\downarrow2.44)$}}}$ & 5.63$_{\textcolor{customred}{\textbf{$(\uparrow0.01)$}}}$ & 10.34$_{\textcolor{customred}{\textbf{$(\uparrow0.78)$}}}$ \\
        GooMBA & \multicolumn{1}{c}{-} & \multicolumn{1}{c}{-} & \multicolumn{1}{c}{-} & \multicolumn{1}{c|}{-} & \multicolumn{1}{c}{-} & \multicolumn{1}{c}{-} & \multicolumn{1}{c}{-} & \multicolumn{1}{c}{-} \\
        \midrule
         & \multicolumn{4}{c|}{\textit{\textbf{Control Flow Flattening (FLA)}}} & \multicolumn{4}{c}{\textit{\textbf{Mixed Boolean-Arithmetic Expression (MBA)}}} \\
        \midrule
        Obf.Pseudocode & \multicolumn{1}{c}{\cellcolor{gray!30}27.10 (\%)} & \multicolumn{1}{c}{\cellcolor{gray!30}68.22 (\%)} & \multicolumn{1}{c}{\cellcolor{gray!30}5.36} & \multicolumn{1}{c}{\cellcolor{gray!30}15.98 ($\times 10^4$)} & \multicolumn{1}{c}{\cellcolor{gray!30}45.38 (\%)} & \multicolumn{1}{c}{\cellcolor{gray!30}73.68 (\%)} & \multicolumn{1}{c}{\cellcolor{gray!30}5.58} & \multicolumn{1}{c}{\cellcolor{gray!30}49.83 ($\times 10^4$)} \\
        \midrule
        Qwen2.5-Coder & 29.44$_{\textcolor{customred}{\textbf{$(\uparrow2.34)$}}}$ & 67.28$_{\textcolor{customgreen}{\textbf{$(\downarrow0.94)$}}}$ & 5.32$_{\textcolor{customgreen}{\textbf{$(\downarrow0.04)$}}}$ & 13.36$_{\textcolor{customgreen}{\textbf{$(\downarrow2.62)$}}}$ & 53.11$_{\textcolor{customred}{\textbf{$(\uparrow7.73)$}}}$ & 76.16$_{\textcolor{customred}{\textbf{$(\uparrow2.48)$}}}$ & 5.55$_{\textcolor{customgreen}{\textbf{$(\downarrow0.03)$}}}$ & 13.87$_{\textcolor{customgreen}{\textbf{$(\downarrow35.96)$}}}$ \\
        Qwen3 & 27.24$_{\textcolor{customred}{\textbf{$(\uparrow0.14)$}}}$ & 68.07$_{\textcolor{customgreen}{\textbf{$(\downarrow0.15)$}}}$ & 5.35$_{\textcolor{customgreen}{\textbf{$(\downarrow0.01)$}}}$ & 15.20$_{\textcolor{customgreen}{\textbf{$(\downarrow0.78)$}}}$ & 50.91$_{\textcolor{customred}{\textbf{$(\uparrow5.53)$}}}$ & \underline{78.13}$_{\textcolor{customred}{\textbf{$(\uparrow4.45)$}}}$ & 5.57$_{\textcolor{customgreen}{\textbf{$(\downarrow0.01)$}}}$ & 17.36$_{\textcolor{customgreen}{\textbf{$(\downarrow32.47)$}}}$ \\
        Codellama & 34.32$_{\textcolor{customred}{\textbf{$(\uparrow7.22)$}}}$ & 66.16$_{\textcolor{customgreen}{\textbf{$(\downarrow2.06)$}}}$ & 5.25$_{\textcolor{customgreen}{\textbf{$(\downarrow0.11)$}}}$ & \hspace{1ex}8.86$_{\textcolor{customgreen}{\textbf{$(\downarrow7.12)$}}}$ & 44.63$_{\textcolor{customgreen}{\textbf{$(\downarrow0.75)$}}}$ & 66.89$_{\textcolor{customgreen}{\textbf{$(\downarrow6.79)$}}}$ & 5.38$_{\textcolor{customgreen}{\textbf{$(\downarrow0.20)$}}}$ & \hspace{1ex}\underline{7.63}$_{\textcolor{customgreen}{\textbf{$(\downarrow42.20)$}}}$ \\
        Llama-3.1 & 21.79$_{\textcolor{customgreen}{\textbf{$(\downarrow5.31)$}}}$ & 67.27$_{\textcolor{customgreen}{\textbf{$(\downarrow0.95)$}}}$ & 5.13$_{\textcolor{customgreen}{\textbf{$(\downarrow0.23)$}}}$ & 12.51$_{\textcolor{customgreen}{\textbf{$(\downarrow3.47)$}}}$ & 36.82$_{\textcolor{customgreen}{\textbf{$(\downarrow8.56)$}}}$ & 70.87$_{\textcolor{customgreen}{\textbf{$(\downarrow2.81)$}}}$ & 5.37$_{\textcolor{customgreen}{\textbf{$(\downarrow0.21)$}}}$ & 15.03$_{\textcolor{customgreen}{\textbf{$(\downarrow34.80)$}}}$ \\
        DeepSeek-V3 & 27.96$_{\textcolor{customred}{\textbf{$(\uparrow0.86)$}}}$ & 62.26$_{\textcolor{customgreen}{\textbf{$(\downarrow5.96)$}}}$ & 5.15$_{\textcolor{customgreen}{\textbf{$(\downarrow0.21)$}}}$ & 10.47$_{\textcolor{customgreen}{\textbf{$(\downarrow5.51)$}}}$ & 49.04$_{\textcolor{customred}{\textbf{$(\uparrow3.66)$}}}$ & 74.19$_{\textcolor{customred}{\textbf{$(\uparrow0.51)$}}}$ & \underline{5.26}$_{\textcolor{customgreen}{\textbf{$(\downarrow0.32)$}}}$ & \hspace{1ex}9.09$_{\textcolor{customgreen}{\textbf{$(\downarrow40.73)$}}}$ \\
        GPT-4o & 26.69$_{\textcolor{customgreen}{\textbf{$(\downarrow0.41)$}}}$ & \underline{78.16}$_{\textcolor{customred}{\textbf{$(\uparrow9.94)$}}}$ & 5.25$_{\textcolor{customgreen}{\textbf{$(\downarrow0.11)$}}}$ & 10.54$_{\textcolor{customgreen}{\textbf{$(\downarrow5.44)$}}}$ & 43.97$_{\textcolor{customgreen}{\textbf{$(\downarrow1.41)$}}}$ & 78.02$_{\textcolor{customred}{\textbf{$(\uparrow4.34)$}}}$ & 5.51$_{\textcolor{customgreen}{\textbf{$(\downarrow0.07)$}}}$ & 10.05$_{\textcolor{customgreen}{\textbf{$(\downarrow39.78)$}}}$ \\
        DeepSeek-R1 & 45.99$_{\textcolor{customred}{\textbf{$(\uparrow18.89)$}}}$ & 72.17$_{\textcolor{customred}{\textbf{$(\uparrow3.95)$}}}$ & 5.27$_{\textcolor{customgreen}{\textbf{$(\downarrow0.09)$}}}$ & \hspace{1ex}\underline{8.78}$_{\textcolor{customgreen}{\textbf{$(\downarrow7.20)$}}}$ & 49.69$_{\textcolor{customred}{\textbf{$(\uparrow4.31)$}}}$ & 74.69$_{\textcolor{customred}{\textbf{$(\uparrow1.01)$}}}$ & 5.39$_{\textcolor{customgreen}{\textbf{$(\downarrow0.19)$}}}$ & 10.26$_{\textcolor{customgreen}{\textbf{$(\downarrow39.57)$}}}$ \\
        OpenAI-o1 & 36.65$_{\textcolor{customred}{\textbf{$(\uparrow9.55)$}}}$ & 65.38$_{\textcolor{customgreen}{\textbf{$(\downarrow2.84)$}}}$ & \underline{5.06}$_{\textcolor{customgreen}{\textbf{$(\downarrow0.30)$}}}$ & 10.06$_{\textcolor{customgreen}{\textbf{$(\downarrow5.92)$}}}$ & 43.67$_{\textcolor{customgreen}{\textbf{$(\downarrow1.71)$}}}$ & 73.10$_{\textcolor{customgreen}{\textbf{$(\downarrow0.58)$}}}$ & 5.28$_{\textcolor{customgreen}{\textbf{$(\downarrow0.30)$}}}$ & 16.16$_{\textcolor{customgreen}{\textbf{$(\downarrow33.67)$}}}$ \\
        ReCopilot & 26.82$_{\textcolor{customgreen}{\textbf{$(\downarrow0.28)$}}}$ & 73.66$_{\textcolor{customred}{\textbf{$(\uparrow5.44)$}}}$ & 5.33$_{\textcolor{customgreen}{\textbf{$(\downarrow0.03)$}}}$ & 16.14$_{\textcolor{customred}{\textbf{$(\uparrow0.16)$}}}$ & 45.64$_{\textcolor{customred}{\textbf{$(\uparrow0.26)$}}}$ & 75.68$_{\textcolor{customred}{\textbf{$(\uparrow2.00)$}}}$ & 5.60$_{\textcolor{customred}{\textbf{$(\uparrow0.02)$}}}$ & 19.01$_{\textcolor{customgreen}{\textbf{$(\downarrow30.82)$}}}$ \\
        ChatDEOB & 48.90$_{\textcolor{customred}{\textbf{$(\uparrow21.80)$}}}$ & 73.98$_{\textcolor{customred}{\textbf{$(\uparrow5.76)$}}}$ & 5.08$_{\textcolor{customgreen}{\textbf{$(\downarrow0.28)$}}}$ & 13.55$_{\textcolor{customgreen}{\textbf{$(\downarrow2.43)$}}}$ & \underline{60.67}$_{\textcolor{customred}{\textbf{$(\uparrow15.29)$}}}$ & 75.27$_{\textcolor{customred}{\textbf{$(\uparrow1.59)$}}}$ & \underline{5.26}$_{\textcolor{customgreen}{\textbf{$(\downarrow0.32)$}}}$ & 13.29$_{\textcolor{customgreen}{\textbf{$(\downarrow36.54)$}}}$ \\
        \midrule
        D810 & \underline{82.96}$_{\textcolor{customred}{\textbf{$(\uparrow55.86)$}}}$ & 75.23$_{\textcolor{customred}{\textbf{$(\uparrow7.01)$}}}$ & 5.45$_{\textcolor{customred}{\textbf{$(\uparrow0.09)$}}}$ & \hspace{1ex}6.32$_{\textcolor{customgreen}{\textbf{$(\downarrow9.66)$}}}$ & \multicolumn{1}{c}{-} & \multicolumn{1}{c}{-} & \multicolumn{1}{c}{-} & \multicolumn{1}{c}{-} \\
        GooMBA & \multicolumn{1}{c}{-} & \multicolumn{1}{c}{-} & \multicolumn{1}{c}{-} & \multicolumn{1}{c|}{-} & 42.72$_{\textcolor{customgreen}{\textbf{$(\downarrow2.66)$}}}$ & 64.29$_{\textcolor{customgreen}{\textbf{$(\downarrow9.39)$}}}$ & 5.61$_{\textcolor{customred}{\textbf{$(\uparrow0.03)$}}}$ & 21.58$_{\textcolor{customgreen}{\textbf{$(\downarrow28.25)$}}}$ \\
        \midrule
         & \multicolumn{4}{c|}{\textit{\textbf{Opaque Predicate (Opaque)}}} & \multicolumn{4}{c}{\textit{\textbf{Immediate Move (ImmMove)}}} \\
        \midrule
        Obf.Pseudocode & \multicolumn{1}{c}{\cellcolor{gray!30}34.84 (\%)} & \multicolumn{1}{c}{\cellcolor{gray!30}69.97 (\%)} & \multicolumn{1}{c}{\cellcolor{gray!30}5.56} & \multicolumn{1}{c}{\cellcolor{gray!30}52.23 ($\times 10^4$)} & \multicolumn{1}{c}{\cellcolor{gray!30}62.78 (\%)} & \multicolumn{1}{c}{\cellcolor{gray!30}60.82 (\%)} & \multicolumn{1}{c}{\cellcolor{gray!30}5.37} & \multicolumn{1}{c}{\cellcolor{gray!30}7.26 ($\times 10^4$)} \\
        \midrule
        Qwen2.5-Coder & 36.32$_{\textcolor{customred}{\textbf{$(\uparrow1.48)$}}}$ & 67.94$_{\textcolor{customgreen}{\textbf{$(\downarrow2.03)$}}}$ & 5.58$_{\textcolor{customred}{\textbf{$(\uparrow0.02)$}}}$ & 19.43$_{\textcolor{customgreen}{\textbf{$(\downarrow32.80)$}}}$ & 59.39$_{\textcolor{customgreen}{\textbf{$(\downarrow3.39)$}}}$ & 60.13$_{\textcolor{customgreen}{\textbf{$(\downarrow0.69)$}}}$ & 5.34$_{\textcolor{customgreen}{\textbf{$(\downarrow0.03)$}}}$ & \hspace{1ex}6.82$_{\textcolor{customgreen}{\textbf{$(\downarrow0.44)$}}}$ \\
        Qwen3 & 35.12$_{\textcolor{customred}{\textbf{$(\uparrow0.28)$}}}$ & 78.85$_{\textcolor{customred}{\textbf{$(\uparrow8.88)$}}}$ & 5.60$_{\textcolor{customred}{\textbf{$(\uparrow0.04)$}}}$ & 23.46$_{\textcolor{customgreen}{\textbf{$(\downarrow28.77)$}}}$ & 60.84$_{\textcolor{customgreen}{\textbf{$(\downarrow1.94)$}}}$ & 60.42$_{\textcolor{customgreen}{\textbf{$(\downarrow0.40)$}}}$ & 5.35$_{\textcolor{customgreen}{\textbf{$(\downarrow0.02)$}}}$ & \hspace{1ex}7.06$_{\textcolor{customgreen}{\textbf{$(\downarrow0.20)$}}}$ \\
        Codellama & 29.96$_{\textcolor{customgreen}{\textbf{$(\downarrow4.88)$}}}$ & 62.12$_{\textcolor{customgreen}{\textbf{$(\downarrow7.85)$}}}$ & 5.49$_{\textcolor{customgreen}{\textbf{$(\downarrow0.07)$}}}$ & 13.68$_{\textcolor{customgreen}{\textbf{$(\downarrow38.55)$}}}$ & 42.16$_{\textcolor{customgreen}{\textbf{$(\downarrow20.62)$}}}$ & 58.39$_{\textcolor{customgreen}{\textbf{$(\downarrow2.43)$}}}$ & 5.10$_{\textcolor{customgreen}{\textbf{$(\downarrow0.27)$}}}$ & \hspace{1ex}5.16$_{\textcolor{customgreen}{\textbf{$(\downarrow2.10)$}}}$ \\
        Llama-3.1 & 24.31$_{\textcolor{customgreen}{\textbf{$(\downarrow10.53)$}}}$ & 61.68$_{\textcolor{customgreen}{\textbf{$(\downarrow8.29)$}}}$ & 5.29$_{\textcolor{customgreen}{\textbf{$(\downarrow0.27)$}}}$ & 14.41$_{\textcolor{customgreen}{\textbf{$(\downarrow37.82)$}}}$ & 34.88$_{\textcolor{customgreen}{\textbf{$(\downarrow27.90)$}}}$ & 58.28$_{\textcolor{customgreen}{\textbf{$(\downarrow2.54)$}}}$ & 5.03$_{\textcolor{customgreen}{\textbf{$(\downarrow0.34)$}}}$ & \hspace{1ex}4.96$_{\textcolor{customgreen}{\textbf{$(\downarrow2.30)$}}}$ \\
        DeepSeek-V3 & 30.19$_{\textcolor{customgreen}{\textbf{$(\downarrow4.65)$}}}$ & 67.38$_{\textcolor{customgreen}{\textbf{$(\downarrow2.59)$}}}$ & 5.34$_{\textcolor{customgreen}{\textbf{$(\downarrow0.22)$}}}$ & 14.77$_{\textcolor{customgreen}{\textbf{$(\downarrow37.46)$}}}$ & 40.41$_{\textcolor{customgreen}{\textbf{$(\downarrow22.37)$}}}$ & 59.80$_{\textcolor{customgreen}{\textbf{$(\downarrow1.02)$}}}$ & 5.10$_{\textcolor{customgreen}{\textbf{$(\downarrow0.27)$}}}$ & \hspace{1ex}4.89$_{\textcolor{customgreen}{\textbf{$(\downarrow2.37)$}}}$ \\
        GPT-4o & 29.34$_{\textcolor{customgreen}{\textbf{$(\downarrow5.50)$}}}$ & 66.63$_{\textcolor{customgreen}{\textbf{$(\downarrow3.34)$}}}$ & 5.57$_{\textcolor{customred}{\textbf{$(\uparrow0.01)$}}}$ & 16.43$_{\textcolor{customgreen}{\textbf{$(\downarrow35.80)$}}}$ & 43.04$_{\textcolor{customgreen}{\textbf{$(\downarrow19.74)$}}}$ & 56.84$_{\textcolor{customgreen}{\textbf{$(\downarrow3.98)$}}}$ & 5.41$_{\textcolor{customred}{\textbf{$(\uparrow0.04)$}}}$ & \hspace{1ex}7.74$_{\textcolor{customred}{\textbf{$(\uparrow0.48)$}}}$ \\
        DeepSeek-R1 & 32.28$_{\textcolor{customgreen}{\textbf{$(\downarrow2.56)$}}}$ & 63.08$_{\textcolor{customgreen}{\textbf{$(\downarrow6.89)$}}}$ & 5.40$_{\textcolor{customgreen}{\textbf{$(\downarrow0.16)$}}}$ & \underline{12.91}$_{\textcolor{customgreen}{\textbf{$(\downarrow39.32)$}}}$ & 40.41$_{\textcolor{customgreen}{\textbf{$(\downarrow22.37)$}}}$ & 57.38$_{\textcolor{customgreen}{\textbf{$(\downarrow3.44)$}}}$ & 5.08$_{\textcolor{customgreen}{\textbf{$(\downarrow0.29)$}}}$ & \hspace{1ex}5.22$_{\textcolor{customgreen}{\textbf{$(\downarrow2.04)$}}}$ \\
        OpenAI-o1 & 28.13$_{\textcolor{customgreen}{\textbf{$(\downarrow6.71)$}}}$ & 68.29$_{\textcolor{customgreen}{\textbf{$(\downarrow1.68)$}}}$ & 5.31$_{\textcolor{customgreen}{\textbf{$(\downarrow0.25)$}}}$ & 17.50$_{\textcolor{customgreen}{\textbf{$(\downarrow34.73)$}}}$ & 38.87$_{\textcolor{customgreen}{\textbf{$(\downarrow23.91)$}}}$ & 57.75$_{\textcolor{customgreen}{\textbf{$(\downarrow3.07)$}}}$ & \underline{5.01}$_{\textcolor{customgreen}{\textbf{$(\downarrow0.36)$}}}$ & \hspace{1ex}\underline{4.82}$_{\textcolor{customgreen}{\textbf{$(\downarrow2.44)$}}}$ \\
        ReCopilot & 35.42$_{\textcolor{customred}{\textbf{$(\uparrow0.58)$}}}$ & 78.05$_{\textcolor{customred}{\textbf{$(\uparrow8.08)$}}}$ & 5.59$_{\textcolor{customred}{\textbf{$(\uparrow0.03)$}}}$ & 22.42$_{\textcolor{customgreen}{\textbf{$(\downarrow29.81)$}}}$ & 62.74$_{\textcolor{customgreen}{\textbf{$(\downarrow0.04)$}}}$ & 60.69$_{\textcolor{customgreen}{\textbf{$(\downarrow0.13)$}}}$ & 5.36$_{\textcolor{customgreen}{\textbf{$(\downarrow0.01)$}}}$ & \hspace{1ex}7.14$_{\textcolor{customgreen}{\textbf{$(\downarrow0.12)$}}}$ \\
        ChatDEOB & \underline{57.70}$_{\textcolor{customred}{\textbf{$(\uparrow22.86)$}}}$ & \underline{80.12}$_{\textcolor{customred}{\textbf{$(\uparrow10.15)$}}}$ & \underline{5.22}$_{\textcolor{customgreen}{\textbf{$(\downarrow0.34)$}}}$ & 14.19$_{\textcolor{customgreen}{\textbf{$(\downarrow38.04)$}}}$ & \underline{74.31}$_{\textcolor{customred}{\textbf{$(\uparrow11.53)$}}}$ & \underline{84.96}$_{\textcolor{customred}{\textbf{$(\uparrow24.14)$}}}$ & 5.50$_{\textcolor{customred}{\textbf{$(\uparrow0.13)$}}}$ & 10.97$_{\textcolor{customred}{\textbf{$(\uparrow3.71)$}}}$ \\
        \midrule
        D810 & \multicolumn{1}{c}{-} & \multicolumn{1}{c}{-} & \multicolumn{1}{c}{-} & \multicolumn{1}{c|}{-} & \multicolumn{1}{c}{-} & \multicolumn{1}{c}{-} & \multicolumn{1}{c}{-} & \multicolumn{1}{c}{-} \\
        GooMBA & \multicolumn{1}{c}{-} & \multicolumn{1}{c}{-} & \multicolumn{1}{c}{-} & \multicolumn{1}{c|}{-} & \multicolumn{1}{c}{-} & \multicolumn{1}{c}{-} & \multicolumn{1}{c}{-} & \multicolumn{1}{c}{-} \\
        \bottomrule 
    \end{tabular}
    \begin{tablenotes}
        \footnotesize
        \item[1] The delta entropy of unobfuscated pseudocode is 5.61, providing a reference value for comparison.
        \item[2] The halstead complexity of unobfuscated pseudocode serves as a baseline, with a value of 7.75$\times 10^4$. 
    \end{tablenotes}
    \end{threeparttable}
    }
    \vspace{-4.0ex}
    \label{tab:rq1}
\end{table}

\noindent\textbf{Unified Evaluation Principle.} Across all evaluation metrics, we consistently quantify improvement or gains using the obfuscated pseudocode as the baseline. Specifically, we compare the deobfuscated output against the unobfuscated pseudocode and measure the relative change with respect to the corresponding obfuscated input, using each metric. An improvement therefore indicates that deobfuscation moves the code closer to the original program along the target dimension.

\vspace{-2.0ex}
\subsection{RQ1: Overall Performance}
\vspace{-2.0ex}

Due to space constraints, \autoref{tab:rq1} presents only LLM deobfuscation results for the x64 architecture at O0 optimization level, while results for other compilation environments are detailed in RQ3.

\vspace{-0.5ex}
The results suggest that the effectiveness of LLM-based deobfuscation depends less on the number of raw parameters and more on the synergy of reasoning capability and domain-specific expertise. In particular, models equipped with strong internal reasoning mechanisms, such as DeepSeek-R1 and OpenAI-o1, consistently outperform others when handling high-entropy obfuscated code. This advantage is especially pronounced for control-flow–intensive transformations, such as FLA. On this task, DeepSeek-R1 achieves a substantial lexical consistency gain of 18.89\%, a 3.95\% improvement in semantic preservation, and a drastic reduction in Halstead complexity ($\downarrow$7.20$\times$$10^4$), relative to the obfuscated pseudocode. Notably, it also attains the best absolute performance among all evaluated models, outperforming both general-purpose and code-specific LLMs on the FLA transformation. These gains suggest that reasoning-oriented inference, exemplified by the test-time scaling paradigm~\cite{snell2024scaling} adopted in DeepSeek-R1, which supports progressive output and refinement during inference, can help resolve complex obfuscation patterns through multiple intermediate reasoning steps rather than single-pass generation. In contrast, general-purpose models like Llama-3.1 exhibit consistent performance degradation under complex obfuscation such as a 5.31\% drop in lexical consistency for the FLA task. This contrast highlights the limitations of direct non-reasoing inference when confronted with deeply entangled control-flow and logic dependencies.

\vspace{-0.5ex}
Parallel to this reliance on reasoning, our findings challenge the applicability of traditional scaling laws to binary code analysis: increasing parameter count alone does not guarantee better performance; instead, domain-specific expertise often outweighs raw model scale. Code-Specific models consistently outperform larger general-purpose models, likely because they are trained to recognize and preserve code structure and control-flow semantics, rather than treating decompiled pseudocode as unstructured or noisy text. For instance, the 32B Qwen2.5-Coder achieves a 16.14\% increase in lexical consistency on the BCF transformation, outperforming the substantially larger 70B Llama-3.1, which shows clear deviations from the original program logic. This advantage stems from pre-training aligned with program analysis tasks rather than from model scale alone, which is further underscored by the 7B domain expert ReCopilot, whose outputs remain more semantically faithful under obfuscation. In comparison, general-purpose models such as GPT-4o tend to generate more aggressively simplified code through "destructive rewriting", occasionally at the cost of semantic fidelity. In addition, as reflected by the aggregate metrics, the fine-tuned baseline ChatDEOB consistently outperforms the domain-pretrained ReCopilot across most obfuscation transformations. This observation suggests that task-specific SFT is more effective than broad domain pre-training for binary code deobfuscation, as it more directly aligns the model with the mapping between obfuscated and deobfuscated code representations.

\vspace{-0.5ex}
Non-LLM deobfuscation methods including D810 and GooMBA can effectively restore the original code within their specific operational scopes. However, as rule-based approaches, they exhibit two notable limitations: they produce syntactically dense outputs with limited readability improvement, as they rely on fixed pattern-matching rules rather than flexible semantic rewriting. In contrast, LLMs offer superior readability improvement through semantic paraphrasing.

\begin{figure*}[t]
\centering
    \begin{subfigure}[b]{0.5\linewidth}
        \scalebox{1.6}{
            \includegraphics[height=2.56cm]{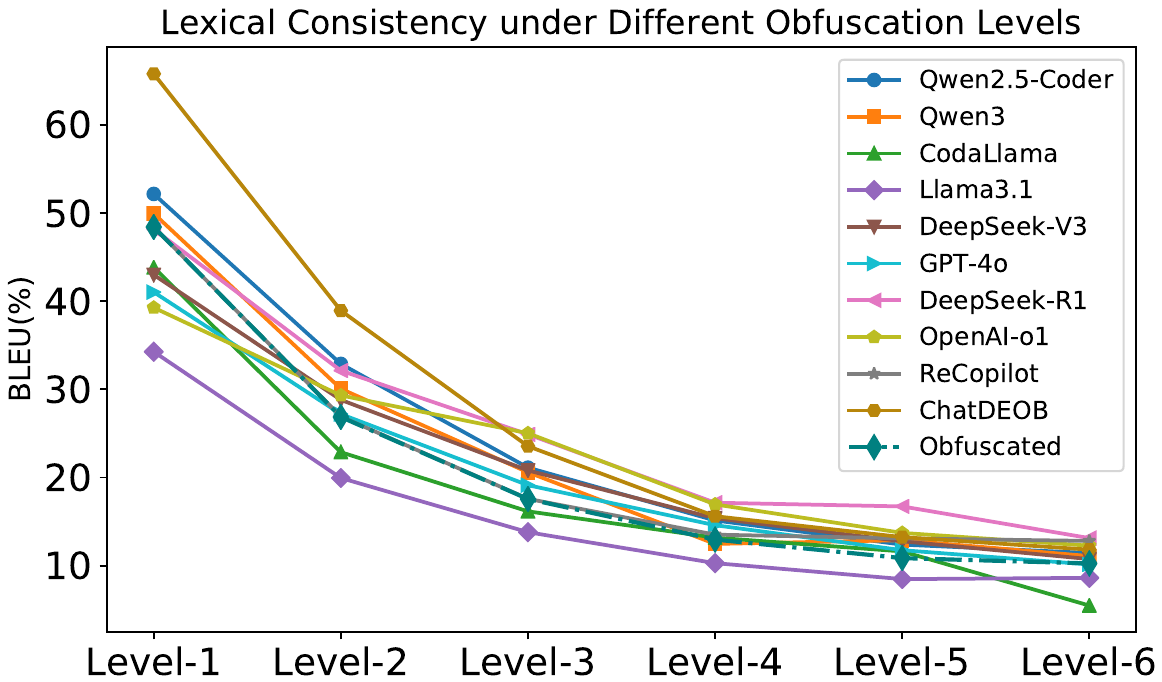}
        }
    \end{subfigure}
    \hspace{-1.3em}  
    \begin{subfigure}[b]{0.5\linewidth}
        \scalebox{1.6}{
            \includegraphics[height=2.56cm]{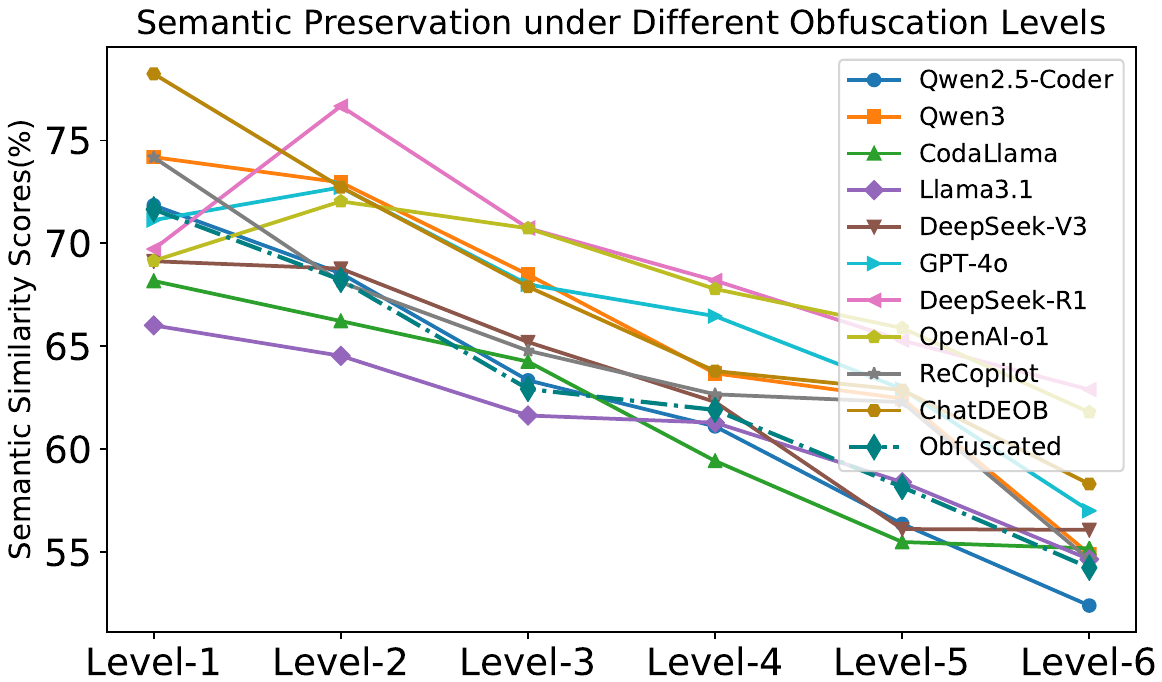}
        }
    \end{subfigure}
    \begin{subfigure}[b]{0.5\linewidth}
        \scalebox{1.6}{
            \includegraphics[height=2.56cm]{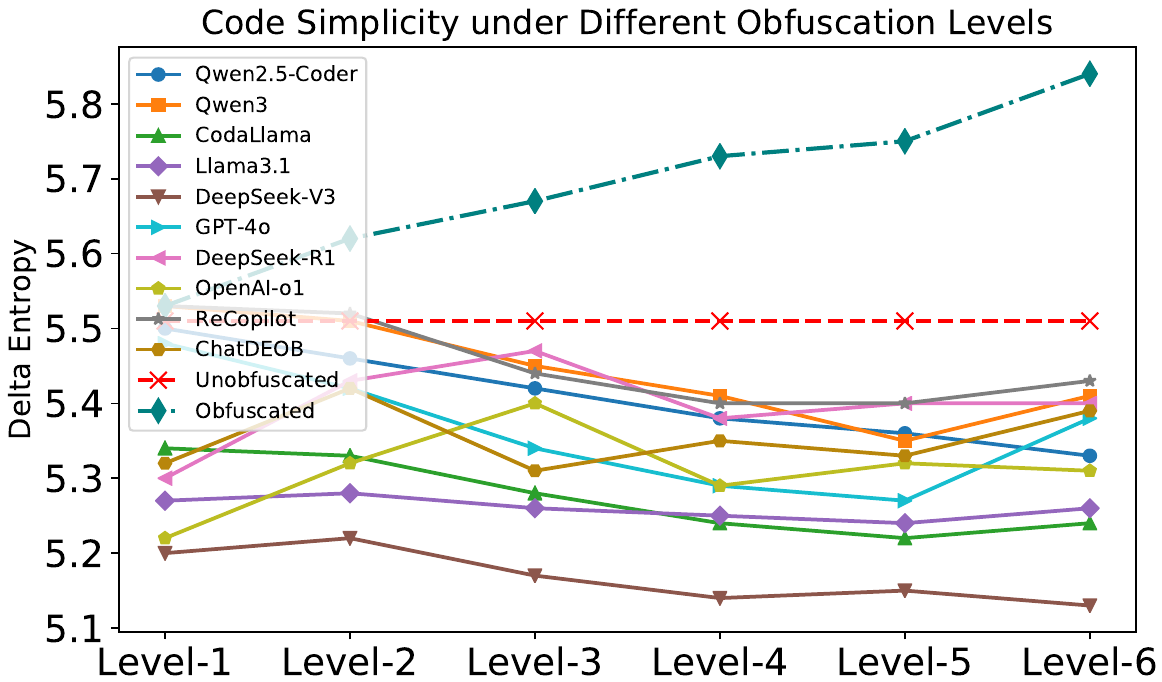}
        }
        \vspace{-1.3ex}  
    \end{subfigure}
    \hspace{-1.3em}  
    \begin{subfigure}[b]{0.5\linewidth}
        \scalebox{1.6}{
            \includegraphics[height=2.56cm]{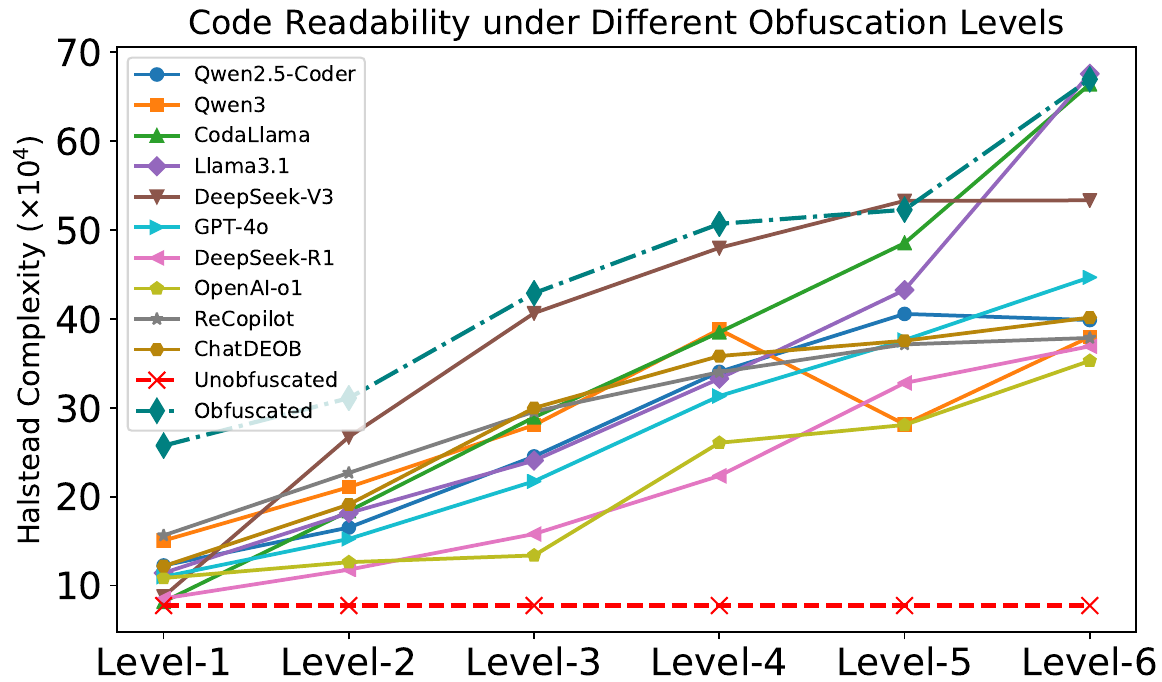}
        }
        \vspace{-1.3ex}  
    \end{subfigure}
    \vspace{1.0ex}
	\caption{Performance under Different Obfuscation Levels.}
    \label{fig:rq2}
    \vspace{-6.0ex}
\end{figure*}

\begin{tcolorbox}[
    colback=gray!5,
    colframe=black,
    arc=0.8mm, auto outer arc,
    boxrule=1pt,
    boxsep=-2pt
]
\textbf{Answering RQ1:} Our evaluation challenges the scaling assumption in binary analysis: increasing parameter count alone does not guarantee better performance; instead, reasoning capability and domain-specific expertise often outweighraw model scale. Additionally, task-specific fine-tuning proves more effective than broad domain pre-training, as it directly aligns the model with the deobfuscation task.
\end{tcolorbox}

\subsection{RQ2: Performance under Different Obfuscation Levels}

To address RQ2, we explore how the combination of obfuscation transformations impacts the overall performance of LLMs, and \autoref{fig:rq2} displays the average results across different combination samplings (detailed in $\S$\ref{dataset construction}). As the non-LLM methods are unable to process combined obfuscation transformations, we exclude their results from our evaluation.

The results across obfuscation levels from Level-1 to Level-6 reveal a non-linear degradation characterized by rapid lexical declines followed by a plateau. Under mild obfuscation, reasoning capabilities offer limited advantages. Instead, ChatDEOB achieves the superior semantic preservation score of 78.24\% at Level-1. Similarly, ReCopilot remains competitive with 74.18\% by capturing pseudocode distribution. However, this advantage diminishes under severe obfuscation. ReCopilot suffers the steepest decline to 54.62\% at Level-6. Even ChatDEOB drops significantly to 58.30\% which reveals the brittleness of standard SFT against compounded transformations. In contrast, reasoning models like DeepSeek-R1 maintain stronger semantic preservation (62.89\%). This gap further highlights the value of multi-step reasoning for handling complex obfuscated logic.

Furthermore, we also observe a divergence between lexical and semantic metrics: as lexical scores collapse, semantic fidelity remains relatively stable. This suggests that LLMs prioritize functional equivalence. They treat deobfuscation as \textit{functional reconstruction} rather than syntactic restoration. Finally, regarding code simplicity, standard models often exhibit inflated complexity arising from repetitive, low-entropy fragments. Conversely, reasoning models leverage deductive capabilities to refactor logic and reduce overall complexity. For instance, DeepSeek-R1 achieves a superior readability score of 36.92$\times10^4$ at Level-6 compared to 40.15$\times10^4$ for ChatDEOB, proving its ability to simplify high-entropy logic effectively.

\begin{tcolorbox}[
    colback=gray!5,
    colframe=black,
    arc=0.8mm, auto outer arc,
    boxrule=1pt,
    boxsep=-2pt
]
\textbf{Answering RQ2:} Deep obfuscation favors reasoning capability over domain expertise. While domain models excel at mild obfuscation, reasoning models demonstrate greater resilience under severe conditions, maintaining semantic fidelity whereas other models fail to resolve the underlying structure.
\end{tcolorbox}

\subsection{RQ3: Performance across Different Compilation Environments}

\begin{figure*}[t]
\centering
    \begin{subfigure}[b]{0.25\linewidth}
        \scalebox{0.98}{
            \includegraphics[height=3.55cm]{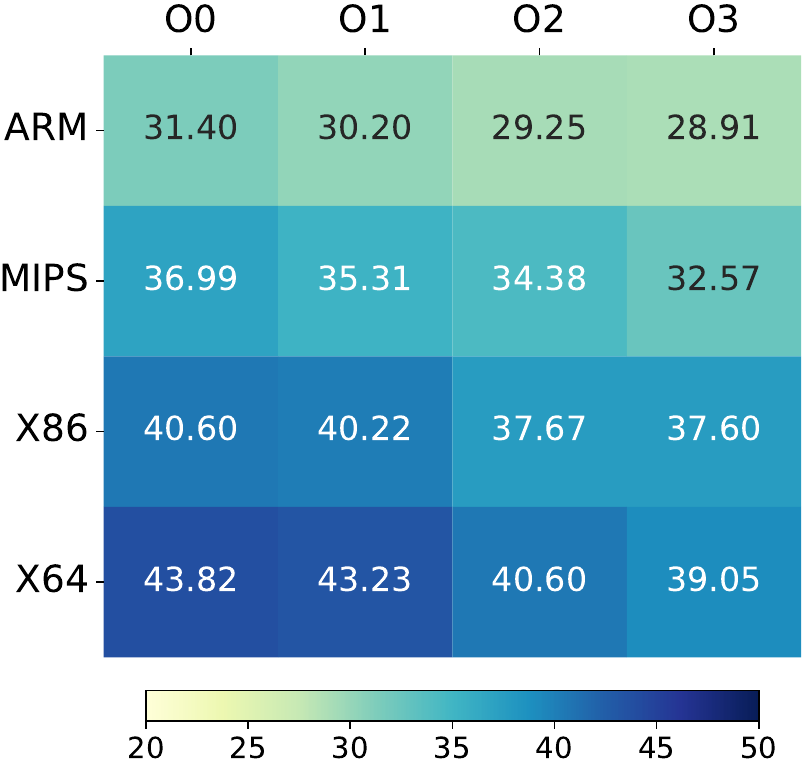}
        }
        \vspace{-1.3ex}  
    \end{subfigure}
    \hspace{-0.4em}  
    \begin{subfigure}[b]{0.25\linewidth}
        \scalebox{0.98}{
            \includegraphics[height=3.55cm]{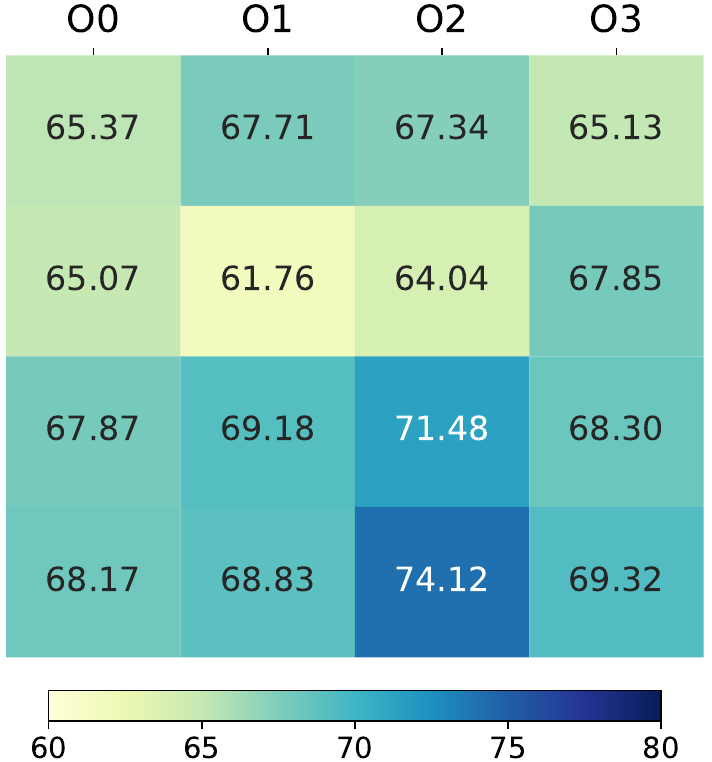}
        }
        \vspace{-1.3ex}  
    \end{subfigure}
    \hspace{-1.5em}  
    \begin{subfigure}[b]{0.25\linewidth}
        \scalebox{0.98}{
            \includegraphics[height=3.55cm]{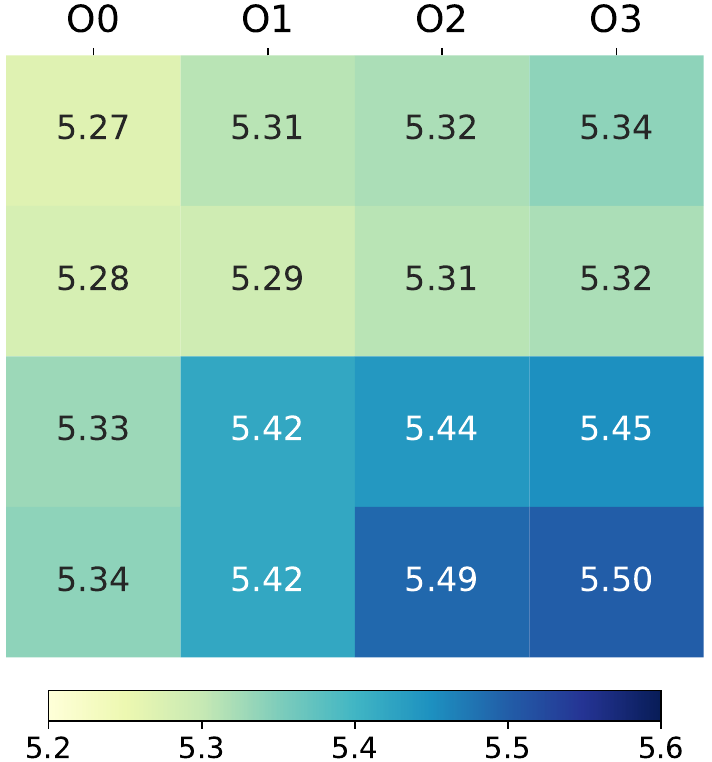}
        }
        \vspace{-1.3ex}  
    \end{subfigure}
    \hspace{-1.5em}  
    \begin{subfigure}[b]{0.25\linewidth}
        \scalebox{0.98}{
            \includegraphics[height=3.55cm]{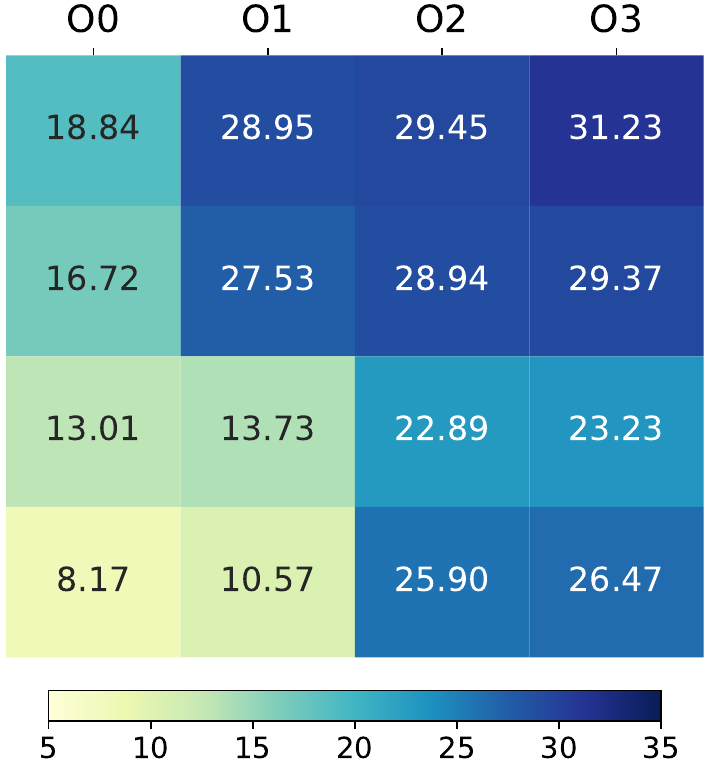}
        }
        \vspace{-1.3ex}  
    \end{subfigure}

    \vspace{1.5ex}  
    \begin{subfigure}[t]{0.25\linewidth}
        \scalebox{0.98}{
            \includegraphics[height=3.35cm]{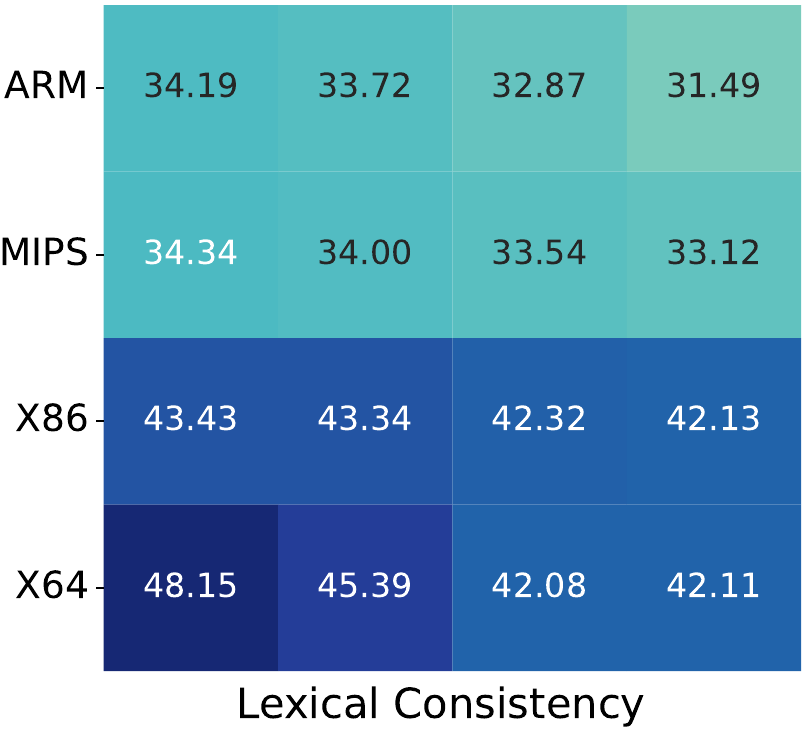}
        }
        \vspace{-1.3ex}  
    \end{subfigure}
    \hspace{-0.4em}  
    \begin{subfigure}[t]{0.25\linewidth}
        \scalebox{0.98}{
            \includegraphics[height=3.35cm]{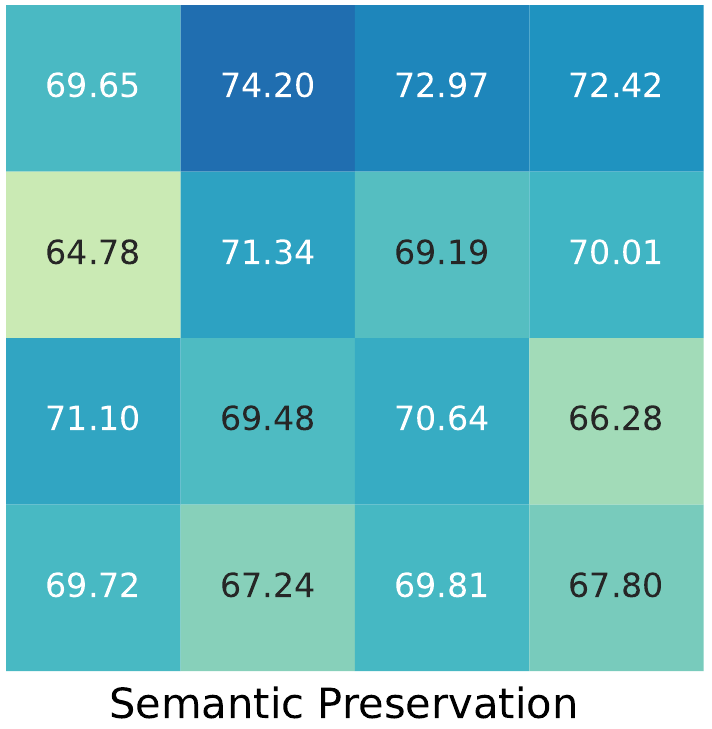}
        }
        \vspace{-1.3ex}  
    \end{subfigure}
    \hspace{-1.5em}  
    \begin{subfigure}[t]{0.25\linewidth}
        \scalebox{0.98}{
            \includegraphics[height=3.35cm]{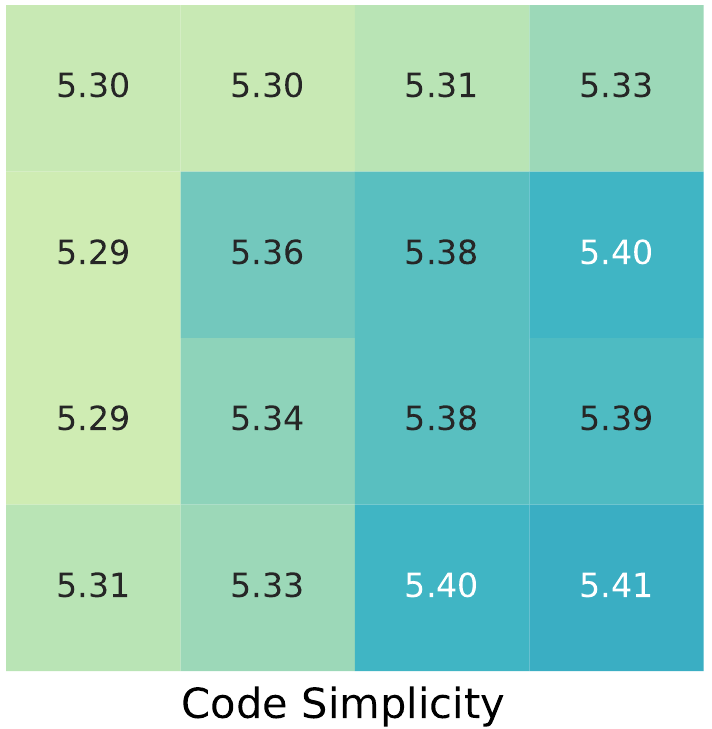}
        }
        \vspace{-1.3ex}  
    \end{subfigure}
    \hspace{-1.5em}  
    \begin{subfigure}[t]{0.25\linewidth}
        \scalebox{0.98}{
            \includegraphics[height=3.35cm]{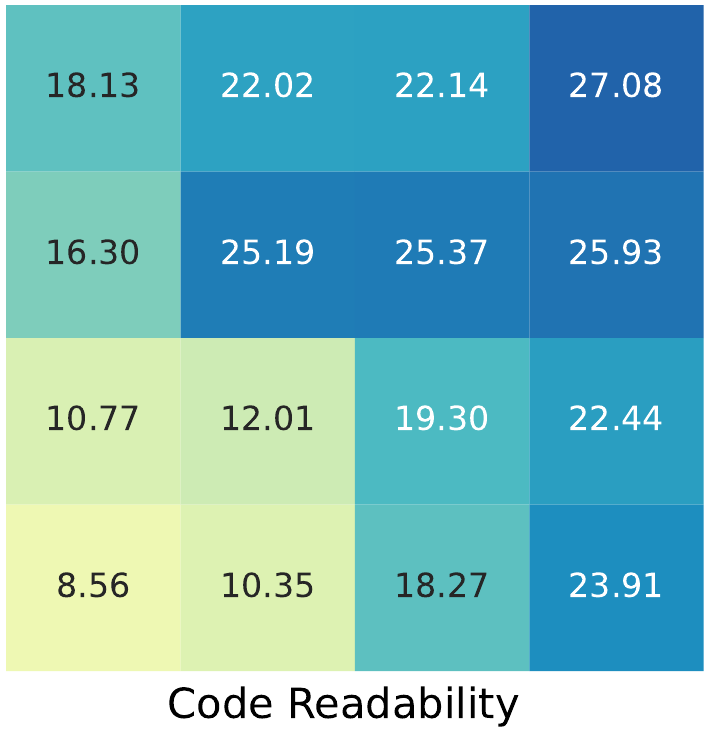}
        }
        \vspace{-1.3ex}  
    \end{subfigure}
	\caption{Performance across Different Architectures and Optimizations. The first row represents the performance of CodeLlama on the four metrics, and the second row shows those of DeepSeek-R1.}
    \label{fig:rq3}
\end{figure*}

To investigate the impact of different architectures and optimization options on LLMs' deobfuscation performance, we conduct an evaluation at Level-1 obfuscation, with average results shown in \autoref{fig:rq3}. Considering experimental costs, we focus on CodeLlama and DeepSeek-R1, which demonstrated strong performance in the previous experiments.

Analysis of performance across architectures reveals a distinct divergence in adaptability between models. CodeLlama exhibits a significant bias towards CISC architectures and its semantic preservation on x64 noticeably outperforms that on RISC architectures like ARM. This likely reflects training data distribution, where x86 and x64 samples are predominant, allowing the model to rely on syntactic familiarity. In contrast, DeepSeek-R1 maintains robust performance independent of the architecture and achieves a semantic score of 72.31\% on ARM. This suggests that strong reasoning capabilities help LLMs generalize beyond architecture-specific patterns, even when facing the verbose pseudocode typical of RISC decompilation.

Furthermore, increasing optimization levels highlight a critical divergence where intensifying severity triggers a universal degradation in surface metrics. This is expected, as aggressive compiler optimizations (e.g., loop unrolling, function inlining) produce convoluted logic that is harder to simplify. However, the models differ in how they handle this complexity. CodeLlama struggles with highly optimized inputs, producing verbose and disorganized outputs (Halstead complexity of 31.23$\times10^4$ on O3-ARM) that mirror the confusion of raw pseudocode. DeepSeek-R1, by contrast, constrains complexity to 27.08$\times10^4$ under identical conditions, effectively refactoring convoluted logic into cleaner code even at peak optimization levels.

\begin{tcolorbox}[
    colback=gray!5,
    colframe=black,
    arc=0.8mm, auto outer arc,
    boxrule=1pt,
    boxsep=-2pt
]
\textbf{Answering RQ3:} Reasoning capabilities help mitigate pre-training biases, enabling robust generalization across CISC and RISC architectures. Additionally, reasoning models handle aggressive compiler optimizations more effectively by refactoring complex logic into cleaner code, whereas other models produce verbose outputs that mirror the confusion of the input.
\end{tcolorbox}

\begin{table}[t]
  \centering
  \renewcommand{\arraystretch}{0.9}
  \caption{Performance of LLMs with In-context Learning on the \sysname.}
  \vspace{1.0ex}
  \setlength{\tabcolsep}{2.5mm}{
  \scalebox{1.0}{
  \begin{threeparttable}
    \begin{tabular}{lccccc}
    \toprule
        \multicolumn{2}{l}{\textbf{Models}} & \textbf{\makecell{Lexical$_{\textcolor{customred}{\textbf{$(\uparrow)$}}}$ \\ Consistency}} & \textbf{\makecell{Semantic$_{\textcolor{customred}{\textbf{$(\uparrow)$}}}$ \\ Preservation}}  & \textbf{\makecell{Code$_{\textcolor{customgreen}{\textbf{$(\downarrow)$}}}$ \\ Simplicity}} & \textbf{\makecell{Code$_{\textcolor{customgreen}{\textbf{$(\downarrow)$}}}$ \\ Readability}} \\
    \midrule
        \multicolumn{2}{l}{Obf.Pseudocode} & \multicolumn{1}{c}{48.40 (\%)} & \multicolumn{1}{c}{71.65 (\%)} & \multicolumn{1}{c}{5.53} & \multicolumn{1}{c}{25.74 ($\times 10^4$)} \\
    \cmidrule{1-6}
        \multirow{4}{*}{CodeLlama}
        & 0-shot & 43.82$_{\textcolor{customgreen}{\textbf{$(\downarrow 4.58)$}}}$ & 68.17$_{\textcolor{customgreen}{\textbf{$(\downarrow 3.48)$}}}$ & \underline{5.34}$_{\textcolor{customgreen}{\textbf{$(\downarrow 0.19)$}}}$ & 8.17$_{\textcolor{customgreen}{\textbf{$(\downarrow 17.57)$}}}$ \\
        & 1-shot & 57.26$_{\textcolor{customred}{\textbf{$(\uparrow 8.86)$}}}$ & 64.92$_{\textcolor{customgreen}{\textbf{$(\downarrow 6.73)$}}}$ & 5.60$_{\textcolor{customred}{\textbf{$(\uparrow 0.07)$}}}$ & 9.41$_{\textcolor{customgreen}{\textbf{$(\downarrow 16.33)$}}}$ \\
        & 3-shot & \hspace{0.8ex}64.81$_{\textcolor{customred}{\textbf{$(\uparrow 16.41)$}}}$ & 71.86$_{\textcolor{customred}{\textbf{$(\uparrow 0.21)$}}}$ & 5.54$_{\textcolor{customred}{\textbf{$(\uparrow 0.01)$}}}$ & \underline{7.03}$_{\textcolor{customgreen}{\textbf{$(\downarrow 18.71)$}}}$ \\
        & 5-shot & \hspace{0.8ex}\underline{66.37}$_{\textcolor{customred}{\textbf{$(\uparrow 21.08)$}}}$ & \underline{72.92}$_{\textcolor{customred}{\textbf{$(\uparrow 1.27)$}}}$ & 5.53$_{\textcolor{customgreen}{\textbf{$(-)$}}}$ \hspace{2.5ex} & 7.73$_{\textcolor{customgreen}{\textbf{$(\downarrow 18.01)$}}}$ \\
    \cmidrule{1-6}
        \multirow{4}{*}{DeepSeek-R1}
        & 0-shot & 48.15$_{\textcolor{customgreen}{\textbf{$(\downarrow 0.25)$}}}$ & 69.72$_{\textcolor{customgreen}{\textbf{$(\downarrow 1.93)$}}}$ & \underline{5.31}$_{\textcolor{customgreen}{\textbf{$(\downarrow 0.22)$}}}$ & 8.56$_{\textcolor{customgreen}{\textbf{$(\downarrow 17.18)$}}}$ \\
        & 1-shot & \hspace{0.8ex}59.72$_{\textcolor{customred}{\textbf{$(\uparrow 11.32)$}}}$ & 68.82$_{\textcolor{customgreen}{\textbf{$(\downarrow 2.83)$}}}$ & 5.45$_{\textcolor{customgreen}{\textbf{$(\downarrow 0.08)$}}}$ & 9.07$_{\textcolor{customgreen}{\textbf{$(\downarrow 16.67)$}}}$ \\
        & 3-shot & \hspace{0.8ex}65.01$_{\textcolor{customred}{\textbf{$(\uparrow 16.61)$}}}$ & \underline{71.02}$_{\textcolor{customgreen}{\textbf{$(\downarrow 0.63)$}}}$ & 5.46$_{\textcolor{customgreen}{\textbf{$(\downarrow 0.07)$}}}$ & 8.41$_{\textcolor{customgreen}{\textbf{$(\downarrow 17.33)$}}}$ \\
        & 5-shot & \hspace{0.8ex}\underline{65.66}$_{\textcolor{customred}{\textbf{$(\uparrow 17.26)$}}}$ & 70.29$_{\textcolor{customgreen}{\textbf{$(\downarrow 1.36)$}}}$ & 5.47$_{\textcolor{customgreen}{\textbf{$(\downarrow 0.06)$}}}$ & \underline{7.82}$_{\textcolor{customgreen}{\textbf{$(\downarrow 17.92)$}}}$ \\
    \bottomrule
    \end{tabular}
    \end{threeparttable}
    }}
  \label{tab:rq4}
\end{table}

\subsection{RQ4: Impact of In-Context Learning}

We investigate the impact of in-context learning on LLM deobfuscation performance. To ensure high-quality reference examples, we manually construct source code and then apply the six obfuscation transformations to it. In this experiment, we focus on CodeLlama and DeepSeek-R1, evaluating four scenarios featuring varying numbers of examples, with the average results shown in \autoref{tab:rq4}. 

The experimental results demonstrate distinct performance trajectories as the number of shots increases. Regarding code complexity, both models consistently exhibit robust performance in code simplicity and code readability across all settings. They significantly reduce the Halstead complexity from the obfuscated baseline of $25.74$ $\times 10^4$ to levels below $10$ $\times 10^4$ and maintain stable delta entropy scores. This suggests that the capability to simplify code and eliminate obfuscation noise is largely intrinsic to the models and less dependent on demonstrations. However, a divergence appears in semantic preservation. CodeLlama displays a positive correlation between context availability and semantic accuracy, peaking at $72.92\%$ in the 5-shot setting, which suggests it effectively leverages pattern matching from examples to reconstruct logic. In contrast, DeepSeek-R1 exhibits diminishing returns in semantic preservation (plateauing at $70.29\%$) despite high lexical consistency. This implies that while few-shot prompts guide the model to mimic the simplified style of the reference code, they may inadvertently interfere with the reasoning-enhanced model's internal reasoning steps, causing it to prioritize surface-level imitation over deep semantic recovery.

\begin{tcolorbox}[
    colback=gray!5,
    colframe=black,
    arc=0.8mm, auto outer arc,
    boxrule=1pt,
    boxsep=-2pt
]
\textbf{Answering RQ4:} In-context learning improves semantic preservation for standard models (e.g., CodeLlama) but yields diminishing returns for reasoning models (e.g., DeepSeek-R1) due to interference with their reasoning steps. Furthermore, the performance on code simplicity and code readability proves to be an intrinsic capability of LLMs, remaining robust and largely independent of the number of demonstrations.
\end{tcolorbox}

\subsection{RQ5: Deobfuscation on Binary Malware}

\begin{figure*}[t]
\centering
    \begin{subfigure}[b]{0.25\linewidth}
        \scalebox{1.0}{
            \includegraphics[height=3.55cm]{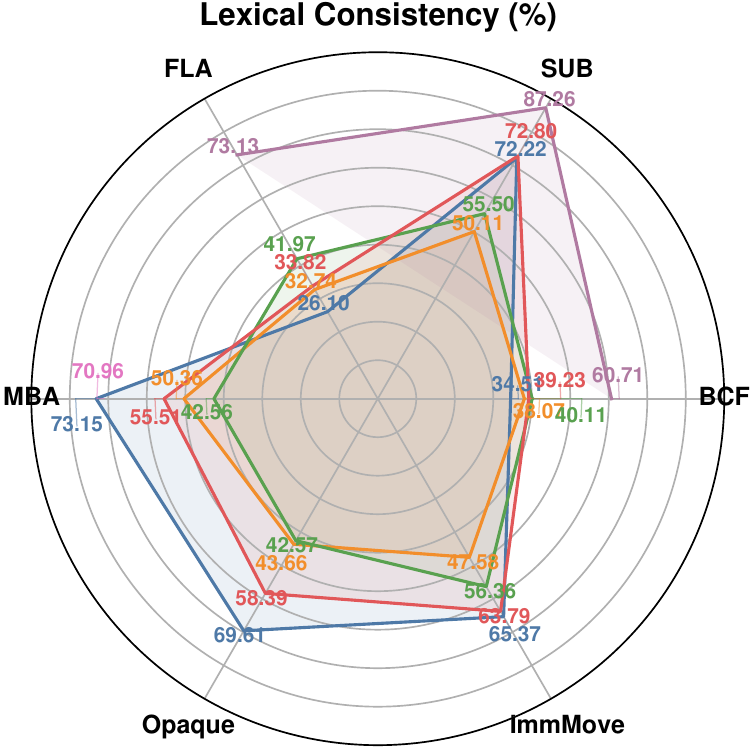}
        }
        \vspace{-1.3ex}  
    \end{subfigure}
    \hspace{-1.2em}  
    \begin{subfigure}[b]{0.25\linewidth}
        \scalebox{1.0}{
            \includegraphics[height=3.55cm]{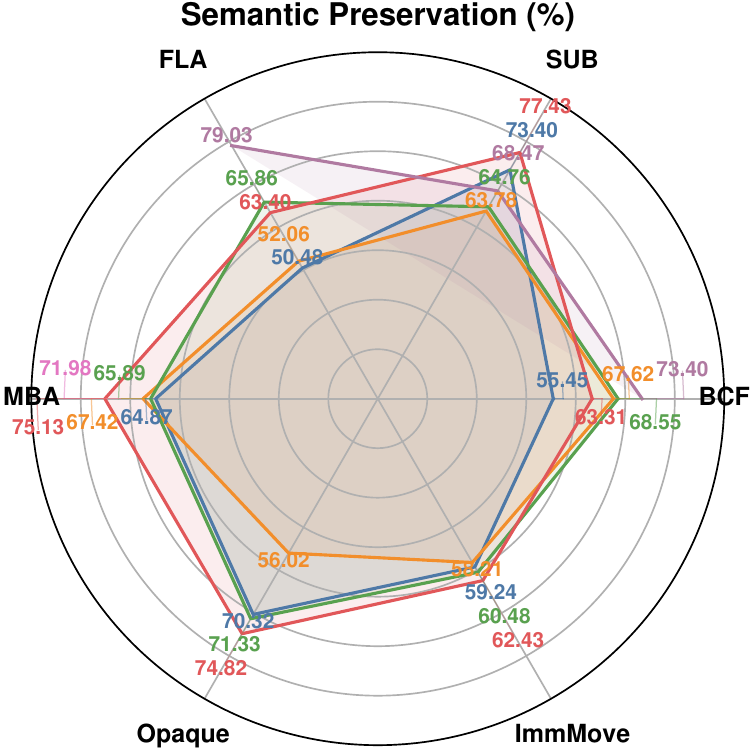}
        }
        \vspace{-1.3ex}  
    \end{subfigure}
    \hspace{-1.2em}  
    \begin{subfigure}[b]{0.25\linewidth}
        \scalebox{1.0}{
            \includegraphics[height=3.55cm]{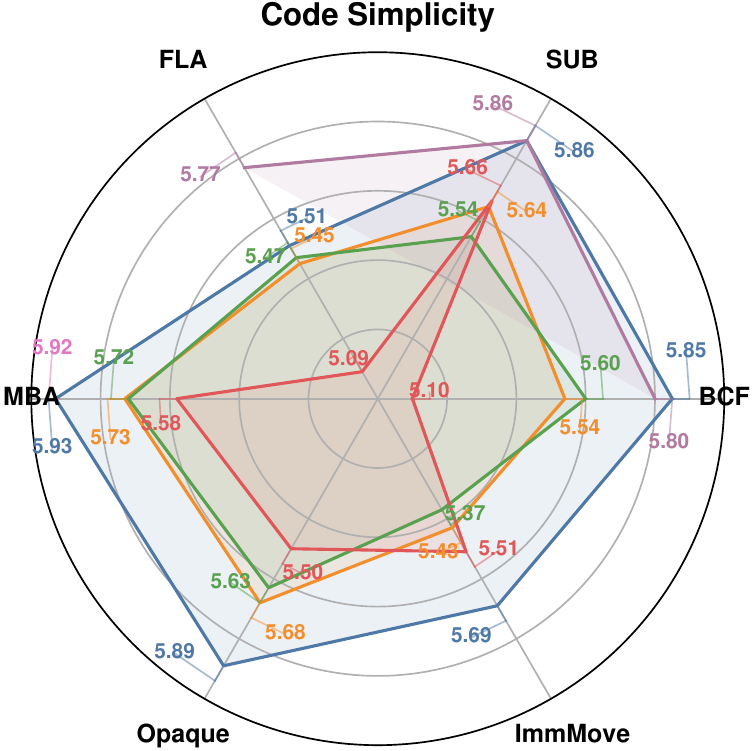}
        }
        \vspace{-1.3ex}  
    \end{subfigure}
    \hspace{-1.2em}  
    \begin{subfigure}[b]{0.25\linewidth}
        \scalebox{1.0}{
            \includegraphics[height=3.55cm]{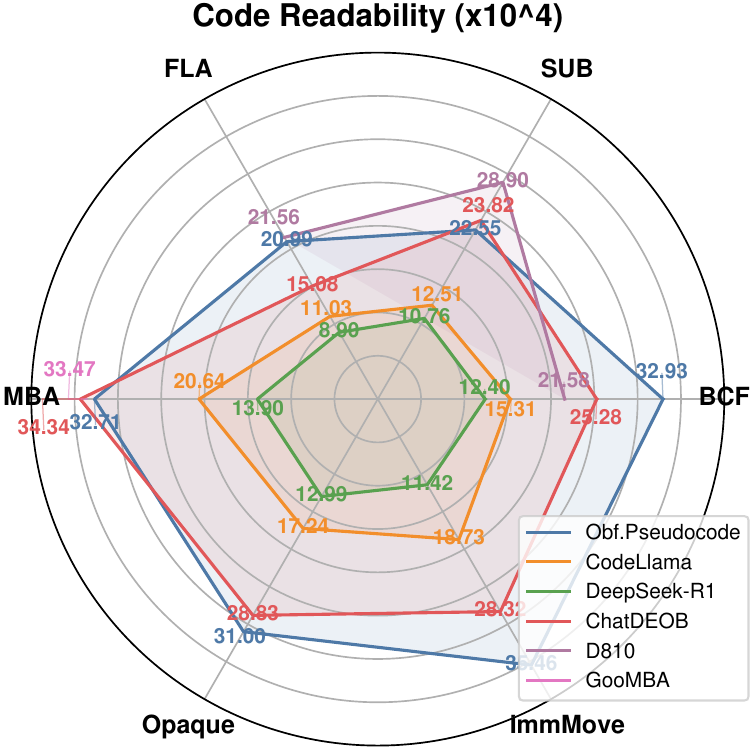}
        }
        \vspace{-1.3ex}  
    \end{subfigure}
    \vspace{1.0ex}
	\caption{Performance of LLMs and Non-LLM Deobfuscation Methods on Malware Dataset.}
    \label{fig:rq5}
    \vspace{-3.0ex}
\end{figure*}

Having demonstrated LLM capabilities on general binary programs, we further investigate their performance on malicious binary programs, a more realistic and challenging scenario. As described in $\S$\ref{dataset construction}, we construct a high-quality evaluation dataset of malicious binaries through manual collection and obfuscation. The comparative performance of LLMs (CodeLlama and DeepSeek-R1) and non-LLM deobfuscation methods is summarized in \autoref{fig:rq5}.

Unlike benign code, realistic malware incorporates manually injected obfuscation measures, introducing significantly higher entropy and complexity. In this challenging context, traditional rule-based tools such as D810 and GooMBA exhibit severe brittleness on unseen variations, whereas LLMs demonstrate superior generalization. Specifically, the reasoning model DeepSeek-R1 significantly enhances code readability by reducing Halstead complexity by approximately 60\% and effectively simplifies deceptive control flows, though it struggles to maintain semantic preservation against high-entropy arithmetic obfuscations. ChatDEOB bridges this gap through supervised fine-tuning, achieving the highest semantic preservation and lowest entropy while effectively neutralizing anti-analysis strategies and eliminating junk code. These results confirm that while reasoning models offer readability improvements for analysis, task-specific fine-tuning is essential for reliably recovering the computational logic of real-world malicious binaries.

\begin{tcolorbox}[
    colback=gray!5,
    colframe=black,
    arc=0.8mm, auto outer arc,
    boxrule=1pt,
    boxsep=-2pt
]
\textbf{Answering RQ5:} Across malicious binary programs, LLMs demonstrate robust generalization capabilities against adversarial obfuscation. Reasoning models excel at enhancing code readability through structural simplification, while domain-specific fine-tuning is essential for semantic recovery and minimizing entropy to restore computational logic.
\end{tcolorbox}

\section{Discussion}

\subsection{Impact of Code Properties}

\begin{figure*}[t]
\centering
    \begin{subfigure}[b]{0.5\linewidth}
        \scalebox{1.6}{
            \includegraphics[height=2.56cm]{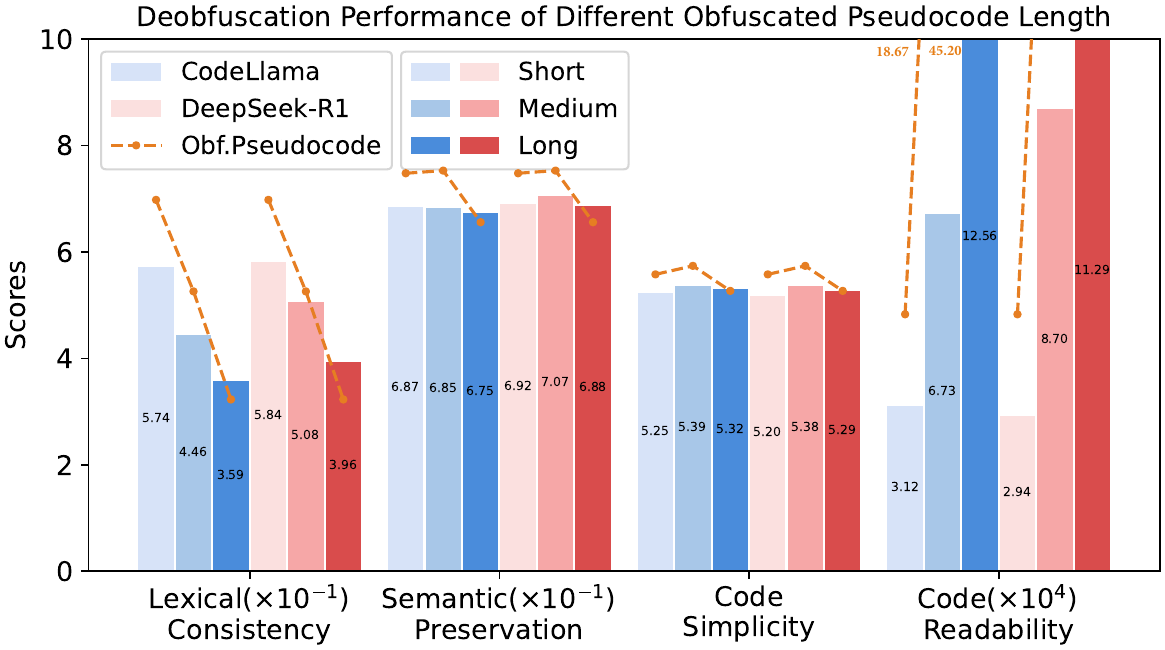}
        }
        \vspace{-1.3ex}  
        \caption{Obfuscated Pseudocode Length}
    \end{subfigure}
    \hspace{-1.3em}  
    \begin{subfigure}[b]{0.5\linewidth}
        \scalebox{1.6}{
            \includegraphics[height=2.56cm]{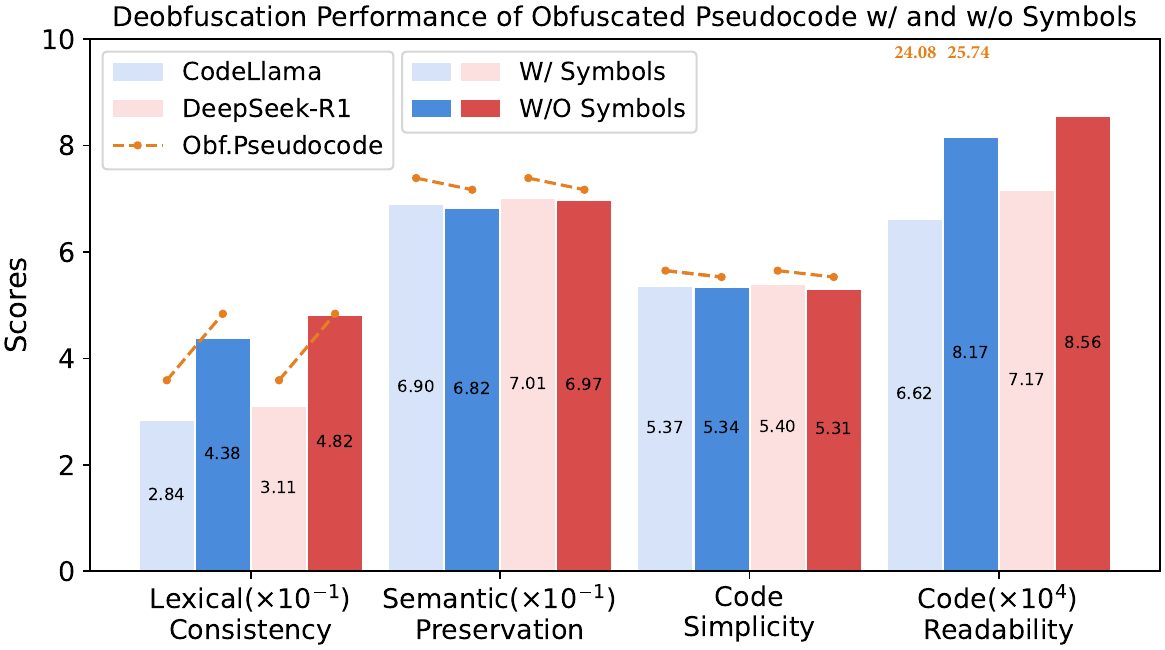}
        }
        \vspace{-1.3ex}  
        \caption{With and Without Symbols}
    \end{subfigure}
	\caption{Deobfuscation Performance with Respect to Pseudocode Length and Symbolic Information.}
    \label{fig:discuss}
    \vspace{-3.0ex}
\end{figure*}

Beyond model types and in-context learning, we also assess how intrinsic code properties affect deobfuscation performance of LLMs. Specifically, we focus on two key attributes: the length of obfuscated pseudocode and the availability of symbolic information. For code length, based on the domain expertise of three senior reverse engineers, we categorize the obfuscated pseudocode functions into three complexity tiers based on Lines of Code (LOC): short (1-50), medium (50-200), and long ($>$200). In our evaluation dataset, the counts for these categories are 564, 1342, and 1094, corresponding to a ratio of approximately 1:2:2. For symbolic information, we consider two scenarios: non-stripped and stripped, and evaluate deobfuscation performance under both settings. The evaluation results are presented in \autoref{fig:discuss}.

The experimental results across varying code lengths show that increasing input size significantly challenges the model's capacity for logic extraction. As input length shifts from Short to Long, the initial Halstead complexity rises to 45.20$\times10^4$, amplifying the logical burden of obfuscation transformations. Under these conditions, DeepSeek-R1 demonstrates a specific advantage in complexity reduction. It successfully lowers the complexity of Long inputs to 11.29$\times10^4$ and outperforms CodeLlama which reduces it to 12.56$\times10^4$. This gap suggests that CodeLlama struggles to filter out the bloated instructions typical of long obfuscated code. In contrast, DeepSeek-R1 effectively identifies and eliminates these redundant logic patterns to produce concise output that is easier to read while maintaining a superior semantic score of 68.8$\%$.

The comparison between inputs with and without symbols reveals that LLMs do not rely on explicit identifiers to understand code logic. Although stripped binaries (W/O Symbols) lack meaningful naming information, DeepSeek-R1 achieves nearly identical semantic preservation scores of 69.7$\%$ without symbols compared to 70.1$\%$ with them. This proves that the model deduces functionality purely from execution patterns rather than relying on identifiers as hints. More notably, we observe that lexical consistency is actually higher without symbols (48.2$\%$) than with them (31.1$\%$). This specific phenomenon occurs because when symbols are present, the model attempts to predict specific identifier names but often fails to match the original source. However, when symbols are stripped, the model defaults to generating standard placeholders and this standardized naming convention aligns more frequently with the ground truth.

\subsection{Lessons Learned and Implications}

\noindent\textbf{Measure what matters - evaluation should capture multiple dimensions.} In this research, we examine deobfuscation performance from four complementary perspectives: lexical consistency, semantic preservation, code simplicity, and code readability. Together, these dimensions provide a balanced view that captures both the accuracy of code recovery and the clarity of its presentation. Such a comprehensive evaluation framework offers a more faithful reflection of deobfuscation quality than relying on a single metric, and it can serve as a foundation for future work in this area.

\noindent\textbf{Potential unlocked - LLMs show promise in binary code deobfuscation.} We conduct a systematic evaluation of LLMs alongside existing deobfuscators using our \sysname and malware dataset. While LLMs may not always outperform non-LLM methods in terms of strict lexical consistency, they demonstrate remarkable capability to generate simplified and readable pseudocode. These results indicate that LLMs are not only capable of deobfuscating binary code but also hold potential to support broader program analysis challenges, such as aiding reverse engineering and identifying security weaknesses.

\noindent\textbf{Power up the engine - boosting LLMs for binary code deobfuscation.} Our experimental analysis demonstrates that employing step-by-step reasoning, leveraging code-specific training, incorporating domain-specific knowledge of binary code, and utilizing in-context learning substantially improve the deobfuscation performance of LLMs. Future research can build on these findings by further strengthening reasoning capabilities, developing specialized training approaches, and refining the selection of contextual examples, paving the way toward more effective binary code deobfuscation and further pushing the boundaries of reverse engineering and software security.

\subsection{Threats to Validity}

\noindent\textbf{External validity.} First, while \sysname constructs the first large-scale dataset of obfuscated binary code from scratch, covering common open-source and commercial obfuscators along with their representative transformations, it cannot encompass all emerging obfuscation techniques. The continuous evolution of obfuscators may introduce transformations not yet represented in our dataset. Second, although we evaluate a diverse range of LLMs, the rapid evolution of the field means we have not exhaustively tested every available model or scale. Our results may vary with the introduction of newer or significantly larger models beyond the scope of this study.

\noindent\textbf{Internal validity.} Our evaluation benchmarks LLMs against reproducible and widely used deobfuscators (e.g., D810 and GooMBA), excluding non-public research prototypes. Consequently, the evaluation may not fully capture the performance of state-of-the-art deobfuscation methods. As research on binary code deobfuscation continues to grow, emerging approaches that become publicly available could serve as valuable baselines for future evaluations. Additionally, as \sysname leverages task-specific SFT strategies for LLMs, variations in training data or fine-tuning techniques might lead to differences in model performance, which were not fully explored in this study.

\noindent\textbf{Construct validity.} While we adopt four complementary metrics to provide a multi-dimensional assessment of deobfuscation quality, thereby reducing the bias of any single metric, the complex nature of binary reverse engineering makes it challenging for any fixed metrics to fully capture the understanding process and cognitive load of human experts. Furthermore, the selected metrics, though comprehensive, may not cover all relevant aspects of binary code analysis, such as the effect of deobfuscation on downstream tasks like vulnerability discovery. Future work could refine these metrics or introduce new ones to better reflect the broader impact of deobfuscation performance.

\section{Conclusion and Future Work}

We present \sysname, the first systematic framework for evaluating LLM capabilities in binary code deobfuscation. Our evaluation shows that reasoning capabilities and domain expertise outweigh raw parameter count for this task. Beyond performance metrics, we observe that LLMs possess an intrinsic capability for code simplification, allowing them to reduce high-entropy noise and normalize logic across varying obfuscation intensities, diverse ISAs, optimization levels, and within malicious binaries. Moreover, our study establishes task-specific SFT as the superior strategy to unlock this potential and surpass rigid rule-baed methods, while in-context learning benefits standard models but yields diminishing returns for reasoning models. As research advances, we believe \sysname will serve as a foundational benchmark, guiding future work to prioritize enhancing internal reasoning processes over model size for complex logic recovery.

In future work, we will continuously introduce additional obfuscation transformations to expand \sysname, enabling evaluation across a broader range of scenarios. Moreover, building on our findings, we will explore more effective LLM-based binary code deobfuscation methods.

\bibliographystyle{unsrt}  
\bibliography{references}

\end{document}